\documentclass[amsmath,preprintnumbers,noshowpacs,twocolumn,plb,superscriptaddress, nofootinbib]{revtex4}
\usepackage{amssymb}
\usepackage{graphicx}
\usepackage{epstopdf}
\usepackage{bm}
\usepackage{epsfig}
\usepackage{graphics}
\usepackage{xspace}
\usepackage{amsfonts}
\usepackage{verbatim}
\usepackage{natbib}
\usepackage{color}
\usepackage{soul}
\usepackage[none]{hyphenat}
\graphicspath{{figures/}}

\begin{document}

\title{Centrality-dependent modification of hadron and jet production \\ in electron-nucleus collisions}

\preprint{CERN-TH-2023-048}
\preprint{LA-UR-23-22860}

\author{Hai Tao Li}
\email{haitao.li@sdu.edu.cn}
\affiliation{School of Physics, Shandong University, Jinan, Shandong 250100, China}
\author{Ze Long Liu}
\email{zelong.liu@cern.ch}
\affiliation{Theoretical Physics Department, CERN, 1211 Geneva 23, Switzerland}
\author{Ivan Vitev}
\email{ivitev@lanl.gov}
\affiliation{Theoretical Division, Los Alamos National Laboratory, Los Alamos, NM, 87545, USA}

\begin{abstract}
Centrality-dependent measurements of hadron and jet cross section attenuation in deep inelastic scattering on nuclei can shed new light on the physics of final-state interactions in the nuclear matter, including the path-length dependence of  the in-medium parton shower formation and evolution. Recent simulation studies have demonstrated the feasibility of experimental centrality determination in $e$A reactions at the electron-ion collider via neutron detection in the zero-degree calorimeter. Motivated by these results, we present the first theoretical calculation of the production rate modification for  hadrons and jets in central and peripheral  $e$Pb collisions. We find that the variation in the suppression of inclusive jet cross section  as a function of centrality is  less than a factor of two. In more differential measurements, such as the distribution of hadrons versus the  hadronization fraction $z_h$, the difference can be enhanced up to an order of magnitude.  
\end{abstract}

\maketitle

\section{Introduction} 
Reactions with nuclei have been an integral part of the study of quantum chromodynamics (QCD) for more than 40 years~\cite{AUBERT1983275}. Cold nuclear matter (CNM) effects in particular have been investigated in electron-nucleus ($e$A)~\cite{Proceedings:2020eah,AbdulKhalek:2021gbh} and proton-nucleus ($p$A) collisions~\cite{Albacete:2013ei,Albacete:2017qng}. These studies include the modification of nuclear structure encoded in parton distribution functions (nPDFs)~\cite{Kovarik:2015cma,Eskola:2016oht,AbdulKhalek:2019mzd}, the non-linear physics of high-gluon densities~\cite{Balitsky:1978ic,McLerran:1993ka,JalilianMarian:1997gr,Kovchegov:1999yj}, and elastic, inelastic and coherent parton scattering in large nuclei~\cite{Wang:2002ri,Qiu:2004qk,Vitev:2007ve,Arratia:2019vju}.

Medium-induced radiative corrections have attracted a lot of attention as a natural mechanism of cross section modification in cold nuclear matter. Specifically, they have been applied  to interpret~\cite{Gavin:1991qk,Arleo:2002ph,Neufeld:2010dz,Xing:2011fb,Arleo:2018zjw,Kang:2015mta}  Drell-Yan and $J/\psi$ suppression at large Feynman-$x$ in minimum bias $p$A~\cite{FNALE772:2000fmo,PhysRevLett.84.3256}, and jet modification in central $p$A at very high energies~\cite{ATLAS:2014cpa,PHENIX:2015fgy}.  Furthermore, in the framework of different theoretical formalisms, including parton energy loss, in-medium evolution, a hybrid approach and renormalization group analysis~\cite{Arleo:2003jz,Chang:2014fba,Li:2020zbk,Ke:2023ixa}, bremsstrahlung from final-state interactions was shown to lead to hadron cross section attenuation in semi-inclusive deep inelastic scattering (SIDIS) on nuclei. The overwhelming majority of these calculations have focused on HERMES collaboration measurements on helium (He), neon (Ne), krypton (Kr) and xenon (Xe)~\cite{Airapetian:2000ks,Airapetian:2003mi,Airapetian:2007vu}, but early EMC collaboration results~\cite{EMC1,ARVIDSON1984381} show the same type of nuclear modification using carbon (C), copper (Cu) and tin (Sn) as targets.

Final-state radiative corrections are not the only possible explanation of HERMES and EMC results. Models on early hadron formation and absorption in nuclear matter have been developed~\cite{Accardi:2002tv,Kopeliovich:2003py} 
and the possibility of universal fragmentation function modification has also been suggested~\cite{Sassot:2009sh,Zurita:2021kli}.   It was found that light hadron measurements at HERMES do not have sufficient discriminating power to uniquely validate or exclude theoretical models~\cite{Accardi:2009qv,Dupre:2011afa}.

The electron-ion collider (EIC) will provide flexible center-of-mass energies and the opportunity to access final states that have not been studied thus far in SIDIS on nuclei. Recently, significant progress has been made in extending the theory of light and heavy hadron suppression~\cite{Li:2020zbk,Ke:2023ixa}, and jet and jet substructure modification~\cite{Li:2020rqj,Li:2021gjw,Devereaux:2023vjz} in $e$A at the EIC. All of these studies have been limited to minimum bias collisions. Centrality-dependent measurements can provide new insights into the physics of final-state interactions in nuclear matter and  centrality class determination has been shown to be feasible via neutron tagging~\cite{Zheng:2014cha,Chang:2022hkt}. 
To this end,  we present  theoretical results on hadron and jet modification in central and peripheral electron-lead ($e$Pb) collisions at the future facility.

The rest of this paper is organized as follows: in Sec.~\ref{sec:theory} we briefly review 
the theoretical formalism for hadron and jet production on protons and nuclei. Discussion of centrality determination in SIDIS and phenomenological results in central and peripheral $e$Pb collisions  are contained in Sec.~\ref{sec:central}. We present our conclusions  in Sec.~\ref{sec:concl}.

\begin{figure}[b!]
	\centering
	\includegraphics[width=0.47\textwidth]{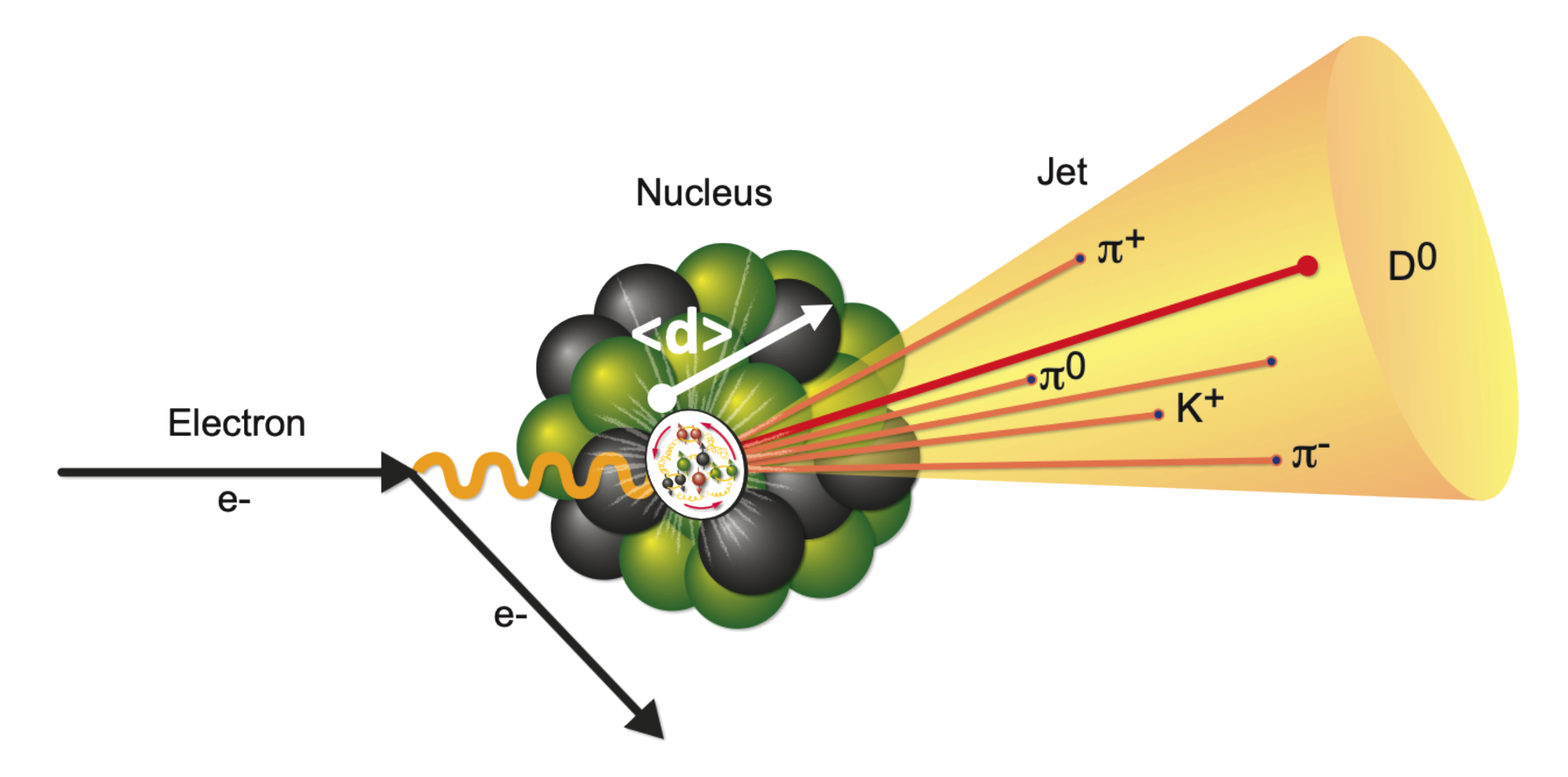}
	\vspace{-0.2cm}
	\caption{Illustration of the concept of centrality in electron-nucleus collisions. The struck quark and the jet initiated by it will see nuclear matter of different mean interaction length $\langle d \rangle$. } 
	\label{fig:Centrality}
\end{figure}

\section{Theoretical formalism}  
\label{sec:theory}

Recent developments in perturbative QCD  have allowed us to place the calculation of semi-inclusive hadron and jet production on the same footing. Using the  formalism of jet functions~\cite{Kang:2016mcy,Dai:2016hzf}, the  collinear differential hadron and jet cross sections can be written in a similar factorized form 
\begin{eqnarray}\label{eq:NLOformh}
 \frac{d \sigma^{\ell N \rightarrow h X}}{dy_h d^{2}{\bf p}_{T,h}} 
&=&\frac{1}{S} \sum_{i, f} \int_{0}^{1} \frac{d x}{x} \int_{0}^{1} \frac{d z}{z^{2}} f^{i / N}(x, \mu) \nonumber \\
& &\times \Big[\hat{\sigma}^{i \rightarrow f}
+f_{\rm ren}^{\gamma /\ell}\left(\frac{-t}{s+u},\mu\right)\hat{\sigma}^{\gamma i \to f}\Big] \nonumber \\
& &\times D^{h / f}(z, \mu)  \, ,  \\
\frac{d \sigma^{\ell N \rightarrow J X}}{dy_J  d^2 {\bf p}_{T,J}} 
&=&\frac{1}{S} \sum_{i, f} \int_{0}^{1} \frac{d x}{x} \int_{0}^{1} \frac{d z}{z^{2}} f^{i / N}(x, \mu) \nonumber \\
& &\times \Big[\hat{\sigma}^{i \rightarrow f}
+f_{\rm ren}^{\gamma /\ell}\left(\frac{-t}{s+u},\mu\right)\hat{\sigma}^{\gamma i \to f}\Big] \nonumber \\ 
& &\times J_{f }(z, p_T R, \mu)   \,  .  \label{eq:NLOformj}
\end{eqnarray}
Here, $ f^{i / N}$ is the parton distribution function (PDF) of parton $i$ carrying a fraction $x$ of in nucleon $N$ momentum.  We denote by $\hat{\sigma}^{i\to f}$ is the lepton-parton scattering cross section producing a final-state parton $f$. The processes that we study as a function of  $p_T$ receive contributions from electron scattering at small angles, where the lepton becomes a source of quasi-real photons.
The corresponding  $\gamma q\to q(g)$, $\gamma q\to g(q)$ and $\gamma g\to q({\bar q})$ processes contribute to the cross section at order $\alpha^2_{\rm em} \alpha_s$  and the Weizs\"acker-Williams (WW) distribution of quasi-real photons is given by a perturbative distribution function $f_{\rm ren}^{\gamma /\ell}\left(y,\mu\right)$~\cite{vonWeizsacker:1934nji,Williams:1934ad}
with $s$, $t$, $u$  the lepton-parton Mandelstam variables.
The analytical results for $\hat{\sigma}^{i\to f}$, $\hat{\sigma}^{\gamma i\to f}$ and $f_{\rm ren}^{\gamma /\ell}\left(y,\mu\right)$  up to ${\cal O}(\alpha_{\rm em}^2 \alpha_s)$ are taken from Ref.~\cite{Hinderer:2015hra}.  $D^{h / f}$ is the standard fragmentation function (FF) from parton $f$ to hadron $h$, taking a momentum fraction $z$. $J_{f}$ is the semi-inclusive jet function (SiJF) initiated by parton $f$. When the jet radius $R$ is small,   logarithms of the type $\ln R$ can be resummed  by  evolving  the  jet  function from the jet scale $p_TR$ to the factorization scale $\mu$.

In $e$A reactions initial-state effects parametrized via nPDFs can alter hadron and jet cross sections.  Our main focus in this paper is the centrality dependence of final-state medium-induced radiative corrections  and we consider observables that minimize or eliminate the cross section modification due to nPDFs. Parton branching in nuclear matter is described by in-medium  splitting kernels  $dN_{ji}^{\mathrm{med} }/{dz d^2\mathbf{k}_{\perp} }$ for the $i \rightarrow j+k$ channel.  We use the results derived in the framework of soft-collinear effective theory with Glauber gluon interaction (SCET$_{\rm G}$)~\cite{Ovanesyan:2011xy,Kang:2016ofv} and verified using a lightcone wavefunction formalism~\cite{Sievert:2018imd,Sievert:2019cwq}.

We calculate numerically the real part of the branching processes, 
\begin{align}
    P_{ j i }^{\mathrm{med},\rm{real}} \left(z, \mathbf{k}_{\perp}\right) = 2\pi\, \mathbf{k}_{\perp}^2  \frac{dN_{j i }^{\mathrm{med}} }{dz d^2\mathbf{k}_{\perp} }  \,,
\end{align}  
for  averaged 
interaction length $\langle d \rangle$ corresponding to different centrality classes, as illustrated in Fig.~\ref{fig:Centrality}. The corresponding 
virtual corrections $P_{ j i }^{\mathrm{med},\rm{vir}}$  are obtained using flavor and momentum sum rules~\cite{Collins:1988wj,Chien:2015vja}. 
Final-state in-medium radiation leads to additional scaling violations~\cite{Altarelli:1977zs} in the fragmentation functions and we implement them in medium-modified  DGLAP 
evolution equations~\cite{Chang:2014fba,Kang:2015mta,Li:2020zbk,Ke:2022gkq}
\begin{eqnarray} \label{eq:fullevol}
    \frac{d {D}^{h/i}\left(x, \mu\right)}{d \ln \mu^{2}}  &=& 
    \sum_{j} \int_{x}^{1} \frac{d z}{z}  
   \left[P_{j i} (z, \mu)   \right. \nonumber   \\  
   && \left. + P_{j i}^{\rm med} (z, \mu)  \right]  {D}^{h/j}\left(\frac{x}{z}, \mu\right)   \, .
\end{eqnarray}
We solve these equations numerically using {\sc Hoppet}~\cite{Salam:2008qg}.

\begin{center}
\begin{table*}[!t]
\begin{tabular}{ |c||c|c|c|c|c|c|c|   } 
 \hline
 Centrality  & 0 -- 1\% &  0 -- 3 \% &  0 -- 10 \% & 60 -- 100 \% & 80 -- 100 \% & 90 -- 100 \% &  0 -- 100 \%   \\ 
 \hline
 $\langle d \rangle [fm]$  &  9.09 &  8.48  &  7.61 & 2.88 & 2.71 & 2.71  &  4.40   \\ 
 $\langle d \rangle / \langle d \rangle_{\rm min. bias}  $    & 2.07 &  1.93   &  1.73  &  0.65 &  0.62 &  0.62 &  1.00   \\ 
 \hline
\end{tabular}
 \caption{Selected centrality classes in $e$Pb collisions at the EIC, the corresponding effective length of cold nuclear matter seen by the scattered parton, and  the ratio relative to the one in minimum bias (0 -- 100 \%) collisions. } 
 \label{table:cent}
 \end{table*}
\end{center}

The SiJFs used to calculate the semi-inclusive jet cross sections also receive medium-induced radiative corrections. We implement them at  next-to-leading order as shown in Refs.~\cite{Kang:2017frl,Li:2018xuv,Li:2020rqj,Li:2021gjw}. The results for quark and gluon initiated jets of transverse momentum $p_T$ and radius parameter $R$  read
\begin{align}\label{eq:sp}
   J_{q}^{\rm{med}}\left(z, p_T R, \mu \right)&=\left[\int_{z(1-z) p_T R}^{\mu} d^2\mathbf{k}_{\perp} f_{q \rightarrow q g}^{\mathrm{med}} \left(z, \mathbf{k}_{\perp}\right)\right]_{+}
    \nonumber \\ 
    &+\int_{z(1-z) p_T R}^{\mu} d^2\mathbf{k}_{\perp} f_{q \rightarrow gq}^{\mathrm{med}} \left(z, \mathbf{k}_{\perp}\right)  \;, 
     \\
    J_{g}^{\rm{med}}\left(z, p_T R, \mu \right)&=
    \nonumber \\ & 
    \hspace{-2cm}
    \left[\int_{z(1-z) p_T R}^{\mu} d^2\mathbf{k}_{\perp} \left(h_{gg} \left(z, \mathbf{k}_{\perp}\right)\left(\frac{z}{1-z}+z(1-z)\right)\right)\right]_{+}
    \nonumber \\ & 
    \hspace{-2cm}
    + n_f \left[ \int_{z(1-z) p_T R}^{\mu} d^2\mathbf{k}_{\perp} f_{g\to q\bar{q}} \left(z, \mathbf{k}_{\perp}\right) \right]_+ 
    \nonumber \\& \hspace{-2cm}
    + \int_{z(1-z) p_T R}^{\mu}d^2\mathbf{k}_{\perp} \Bigg( h_{gg}(x, \mathbf{k}_{\perp})\left(\frac{1-z}{z}+\frac{z(1-z)}{2} \right)
     \nonumber  \\  & 
    + n_f  f_{g\to q\bar{q}}(z,\mathbf{k}_{\perp}) \Bigg) \;,
    \label{eq:sp1} 
\end{align}
where we have denoted  ${dN_{j i }^{\mathrm{med}} }/{dz d^2\mathbf{k}_{\perp} }  \equiv    f_{i\rightarrow jk}^{\mathrm{med}} \left(z, \mathbf{k}_{\perp}\right)$ for brevity.   In Eq.~(\ref{eq:sp1})
\begin{align}
    h_{gg} \left(z, \mathbf{k}_{\perp}\right)  =& \frac{ f_{g \rightarrow gg }^{\mathrm{med}}
   \left(z, \mathbf{k}_{\perp}\right)}{ \frac{z}{1-z} + \frac{1-z}{z}+z(1-z)} \;. 
\end{align}
 In the equations above all,  singularities when $z\to 1$ are regularized by the plus-distribution function that has the standard definition.

\section{Centrality dependent nuclear modification} 
\label{sec:central}

To study the centrality dependent nuclear modification, we are motivated by recent simulations of constraints on nuclear geometry in $e$A reactions using the Monte Carlo event generator BeAGLE~\cite{Chang:2022hkt}. The idea behind this more differential approach is to measure the energy deposited in the zero-degree calorimeter~\cite{Zheng:2014cha} at the EIC and correlate it to collision centrality,  and the effective path length $\langle d \rangle$. A subset of effects, such as shadowing or assumed initial particle formation time, were studied and found to not significantly affect the energy distribution in the ZDC. The correlation between the centrality classes and the energy deposition remains robust when such effects are taken into account in simulation.

With this in mind, the average interaction length  of a parton in a Pb  nucleus as a function of centrality obtained in BeAGLE  is given in Table~\ref{table:cent}. In the top  0-1\% central events $\langle d \rangle $ is twice as large as the one in minimum bias collisions. In the most peripheral    90-100\%  events  $\langle d \rangle $ is almost twice as small as the minimum bias one.  In this paper, we pick two more representative examples of centrality selection -  a central  - 0-10\% class and a peripheral 80-100 \% class.  Next, we calculate grids of in-medium splitting functions~\cite{Ovanesyan:2011xy,Kang:2016ofv,Sievert:2018imd,Sievert:2019cwq} while constraining  nuclear geometry  to  yield  the enhancement or reduction of the average interaction length relative to the minimum bias one  as given in   Table~\ref{table:cent}.

With the numerically evaluated splitting functions at hand and the theoretical framework described in Sec.~II we now turn to phenomenology. In our calculations for the baseline $ep$ collisions  we  use CT14nlo PDF sets~\cite{Dulat:2015mca} 
with the strong coupling constant provided by  {\sc  Lhapdf6}~\cite{Buckley:2014ana}. For the case of semi-inclusive hadron production, fragmentation functions into light pions  are taken directly from the HKNS parameterization in  Ref.~\cite{Hirai:2007cx}.  Heavy quark fragmentation into  $D$- and $B$-mesons  at the scale $\mu = 2 m_Q$ is evaluated perturbatively using heavy quark effective theory (HQET)~\cite{Braaten:1994bz,Cheung:1995ye} and evolved to a higher scale. When we consider reactions with nuclei, such as the $e$Pb case of interest,  we use the nCTEQ15FullNuc PDF sets~\cite{Kovarik:2015cma}.  Consistent with  Ref.~\cite{Li:2020zbk} and more recently Ref.~\cite{Ke:2023ixa},  we fix the gluon transport coefficient in cold nuclear matter $\langle k_\perp^2\rangle/\lambda_g=0.12$ GeV$^2$/fm and 
$\langle k_\perp^2\rangle/\lambda_q=0.053$ GeV$^2$/fm.

\begin{figure*}[t!]
	\centering
	 	\centering
 	\includegraphics[width=0.48\textwidth]{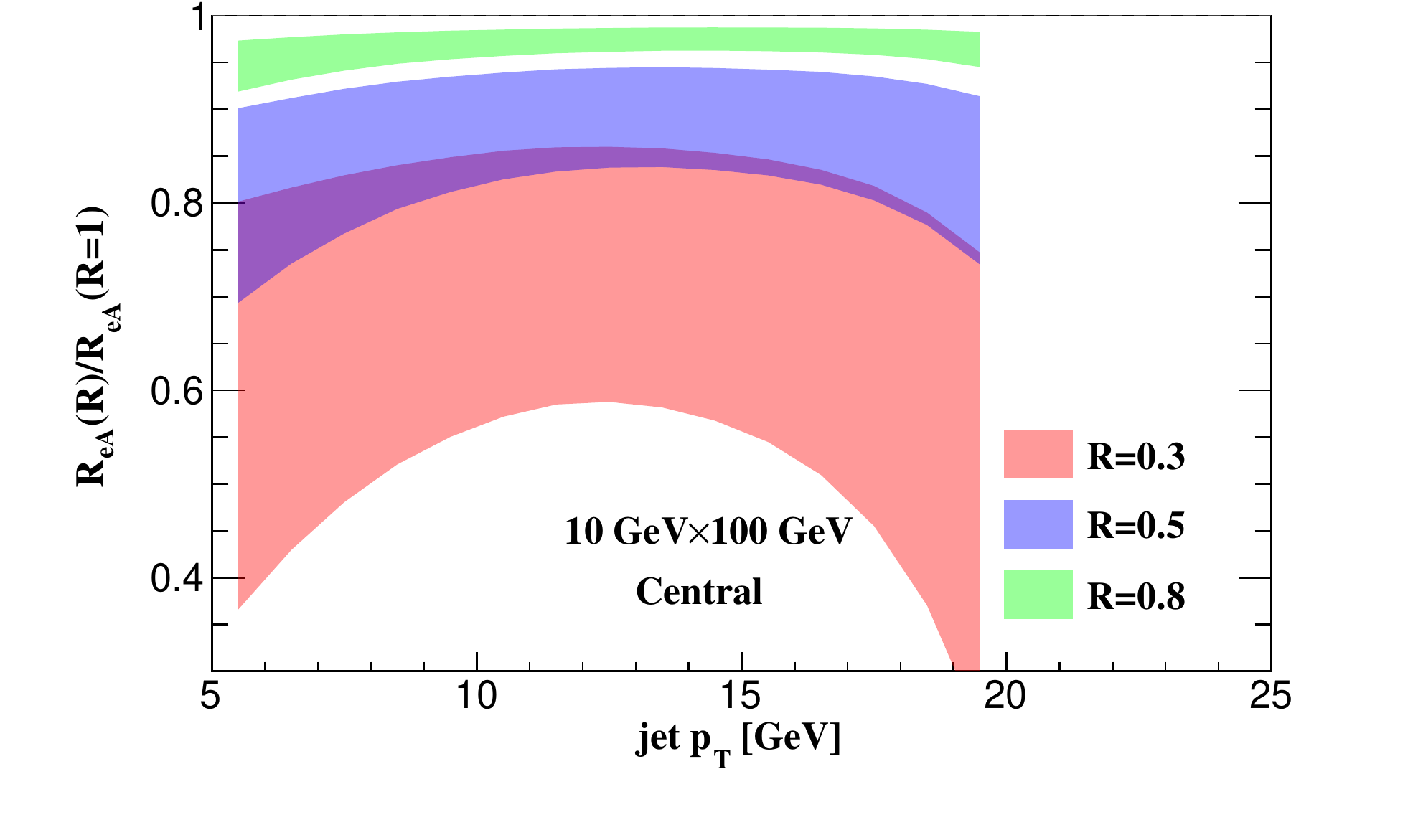}\,\,\,
 	\includegraphics[width=0.48\textwidth]{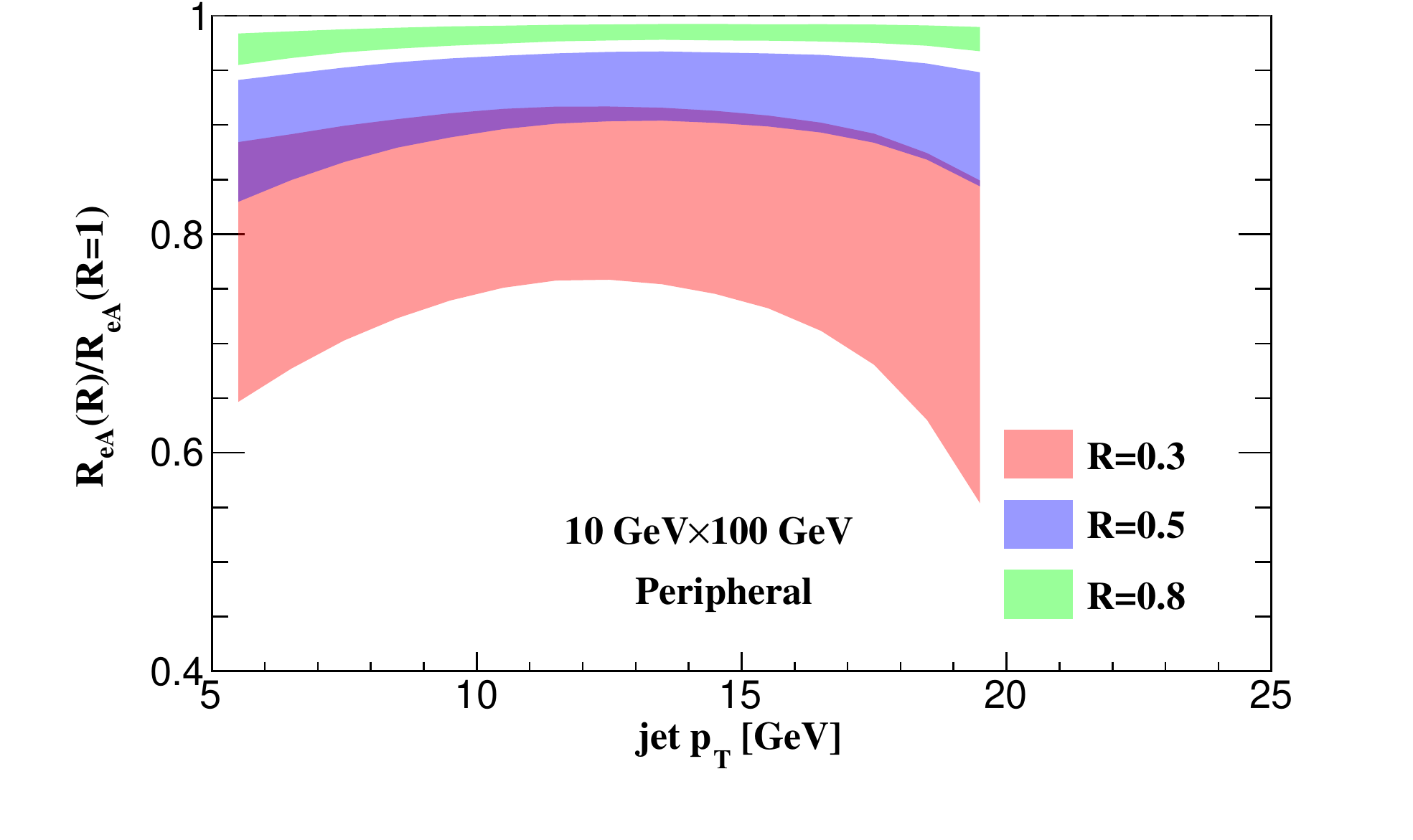}
 	\includegraphics[width=0.48\textwidth]{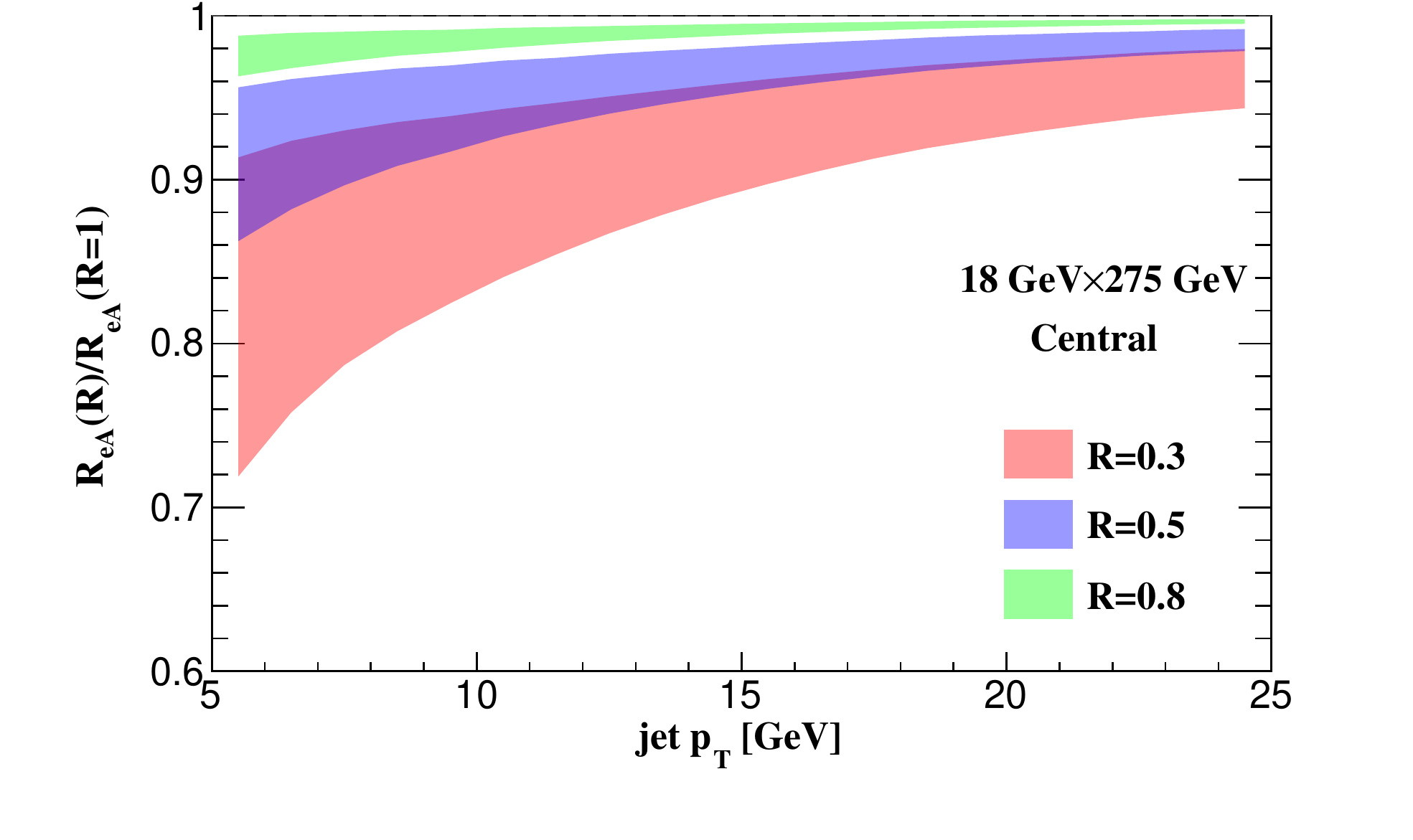}	\,\,\,
 	\includegraphics[width=0.48\textwidth]{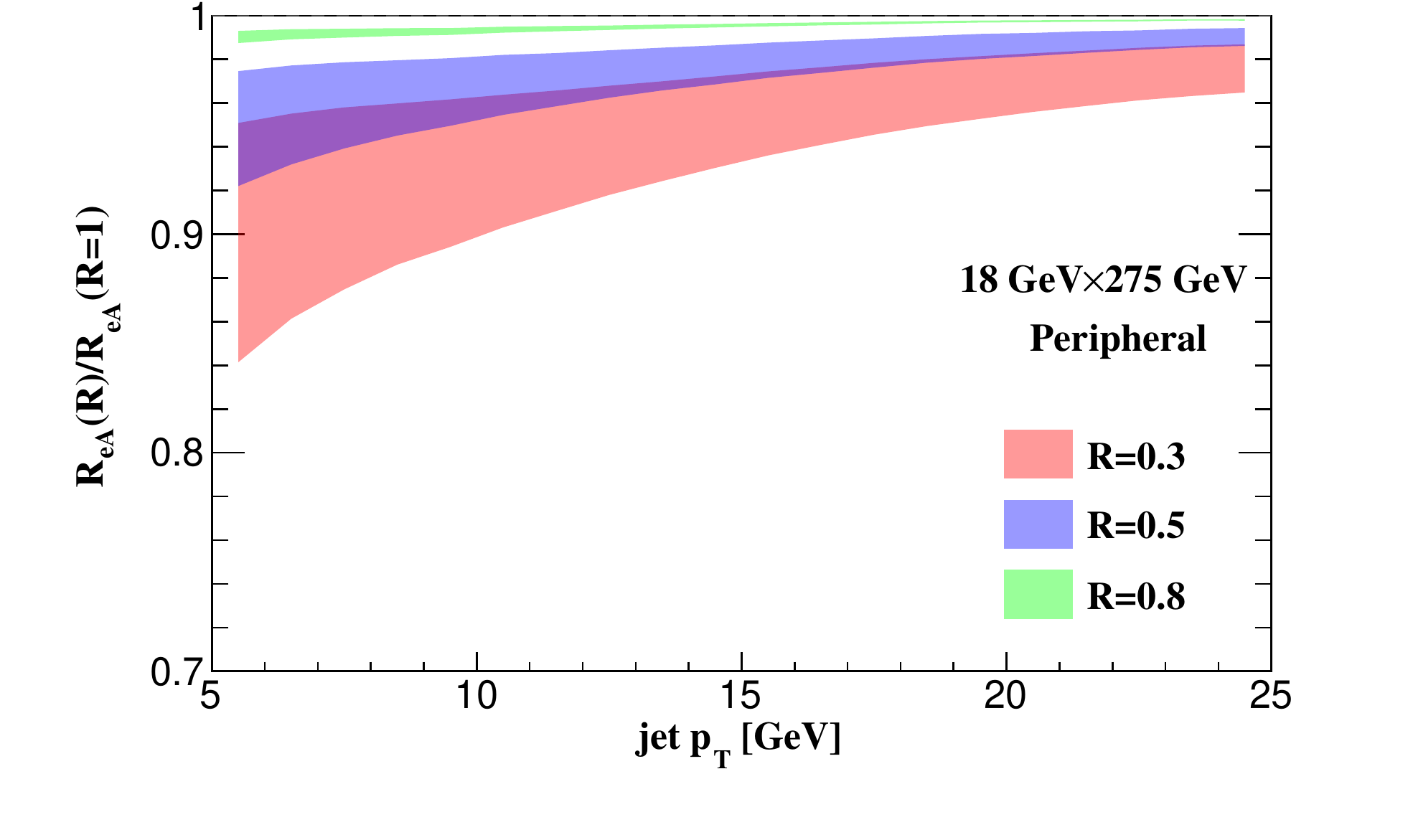}		
	\vspace{-0cm}
	\caption{ Relative modifications of the inclusive jet cross section  $R_{e\rm A}(R)/R_{e\rm A}(R=1)$ for three radius choices $R=0.3, 0.5, 0.8$ in the rapidity interval  $2< \eta < 4$. The upper  panels are for  10 $\times$ 100~GeV  $e$Pb collisions and the bottom panels are for  18 $\times$ 275~GeV  $e$Pb collisions. Central reactions are on the left and peripheral reactions are on the right.  }
	\label{fig:ReAReA1}
\end{figure*}

We first consider jets reconstructed with a radius parameter $R$
and define the  centrality dependent nuclear modification  in electron-nucleus collisions through the ratio 
\begin{align}
    R_{\rm eA}(R) = \frac{1}{\Delta_b T_A(b)}   \frac{\int_{\eta1}^{\eta2} d\sigma/d\eta dp_T\bigskip
    |_{e+A}}{   \int_{\eta1}^{\eta2}  d\sigma/d\eta dp_T\big|_{e+p}}\,.
\label{nuclemod}
\end{align}
Here, the nuclear thickness function at impact parameter $b$ is
\begin{equation}
T_A(b) =  \int_{-\infty}^{\infty}  \rho(z,b) dz  \; , 
\end{equation}
and $\Delta_b = 2\pi b  db$ is the differential area around the impact parameter $b$  such that $\sum_b \Delta_b T_A(b) =A$. In other words,  $R_{e\rm A}(R)$ is the per nucleon cross section modification for the relevant impact parameters corresponding to the centrality class. Earlier work on hadron and jet production in minimum bias $e$A collisions has already provided useful guidance on how to study final-state interactions~\cite{Li:2020rqj,Li:2020zbk,Li:2021gjw}. In particular, they can be separated from initial-state nuclear PDFs~\cite{Kovarik:2015cma,Eskola:2016oht} by taking  the ratio of nuclear modification for a small radius jet to the modification for  a large radius jet    $R_{e\rm A}(R)/ R_{e\rm A}(R=1)$. This strategy works very well, eliminating initial-state effects to less than a few \%~\cite{Li:2020rqj,Li:2021gjw}.

In Fig.~\ref{fig:ReAReA1} we show the double  modification ratio $R_{e\rm A}(R)/ R_{e\rm A}(R=1)$ for three different choices $R=0.3$ (red band), $0.5$ (blue band), and $0.8$ (green band). The bands correspond to varying the cold nuclear matter transport parameters by a factor of two relative to the nominal values quoted above. The idea behind normalizing this observable to the $R_{e\rm A}$  for a large radius jet is that final-state effect  for $R=1$ will be minimal. Even though the medium induced parton shower is broader than the vacuum one, most of it will be contained in a unit radius. Conversely, by choosing smaller radii an increasingly larger fraction of the shower energy will be redistributed outside of the jet cone, leading to cross section suppression. We also choose the forward proton/nucleus going direction $2< \eta < 4$ since the jet energy in this kinematic region is the smallest in the rest frame of the nucleus, leading to larger final-state effects. The top row  of panels shows  10 GeV ($e$) $\times$  GeV (Pb) collision and the bottom row of  panels is for 18 GeV ($e$) $\times$ 275 GeV (Pb) collisions.  On the left we show the 0-10\% centrality selection and the 80-100\% centrality class is on the right.

Our calculations show that the nuclear modification is the largest at relatively small transverse momenta. At the same time, it depends on the steepness of the $p_T$ spectra and the effects become larger again close to the kinematic edges of phase space, as seen in the upper panels of Fig.~\ref{fig:ReAReA1}.  For large radius jets the relative modification    $R_{\rm eA}(R=0.8)/ R_{\rm eA}(R=1)$ is small, $\leq 5\%$. On the other hand, for small radius jets     $R_{\rm eA}(R=0.8)/ R_{\rm eA}(R=1)$ in central $e$Pb reactions can show more than a factor of two suppression.  At higher center of mass energies the modification is smaller, as expected, and decreases monotonically with $p_T$. By comparing the left and right panels of Fig.~\ref{fig:ReAReA1} we see clearly that final-state effects depend on the thickness of nuclear matter.

 \begin{figure}[!t]
    \centering
    \includegraphics[width=0.48\textwidth]{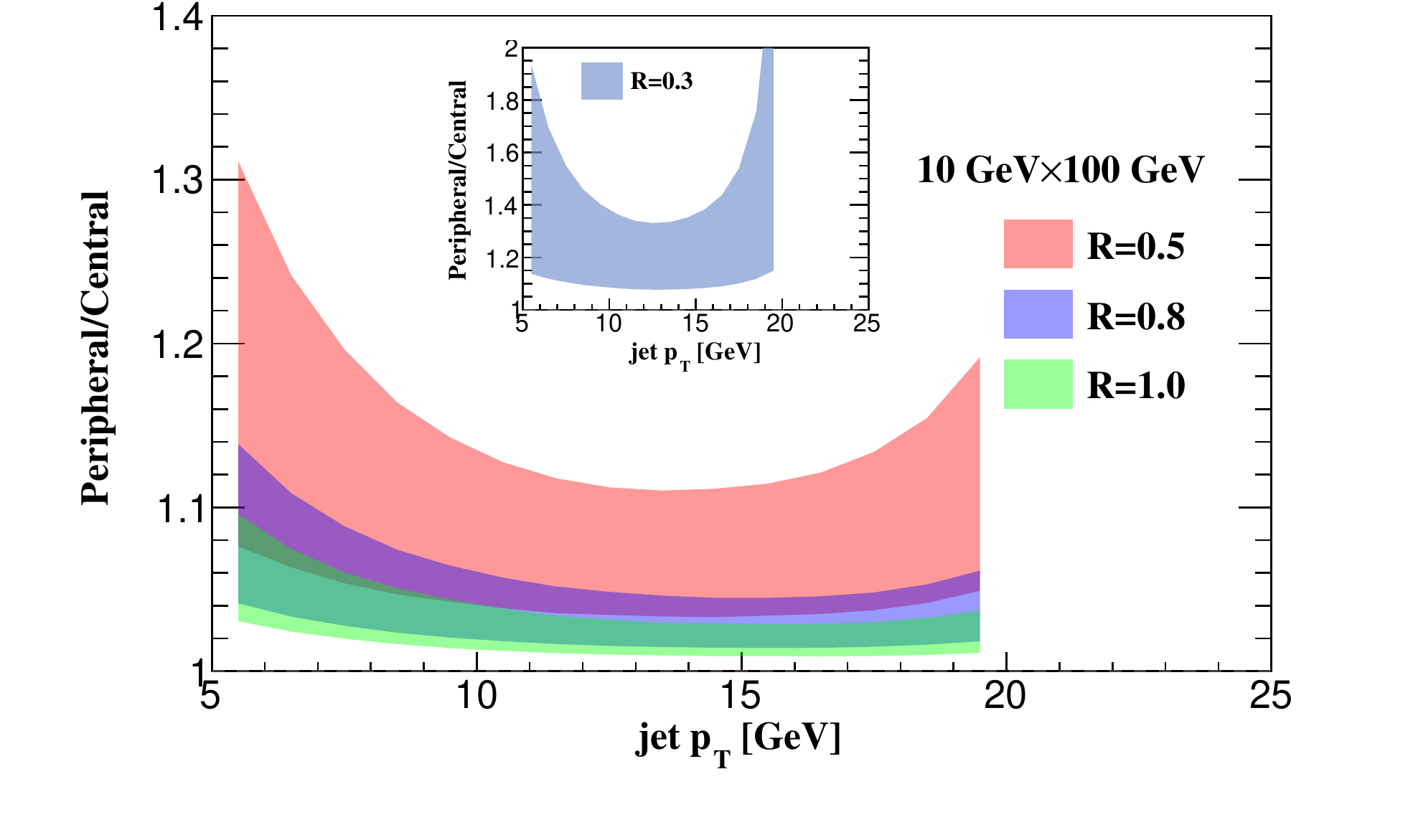}
     \includegraphics[width=0.48\textwidth]{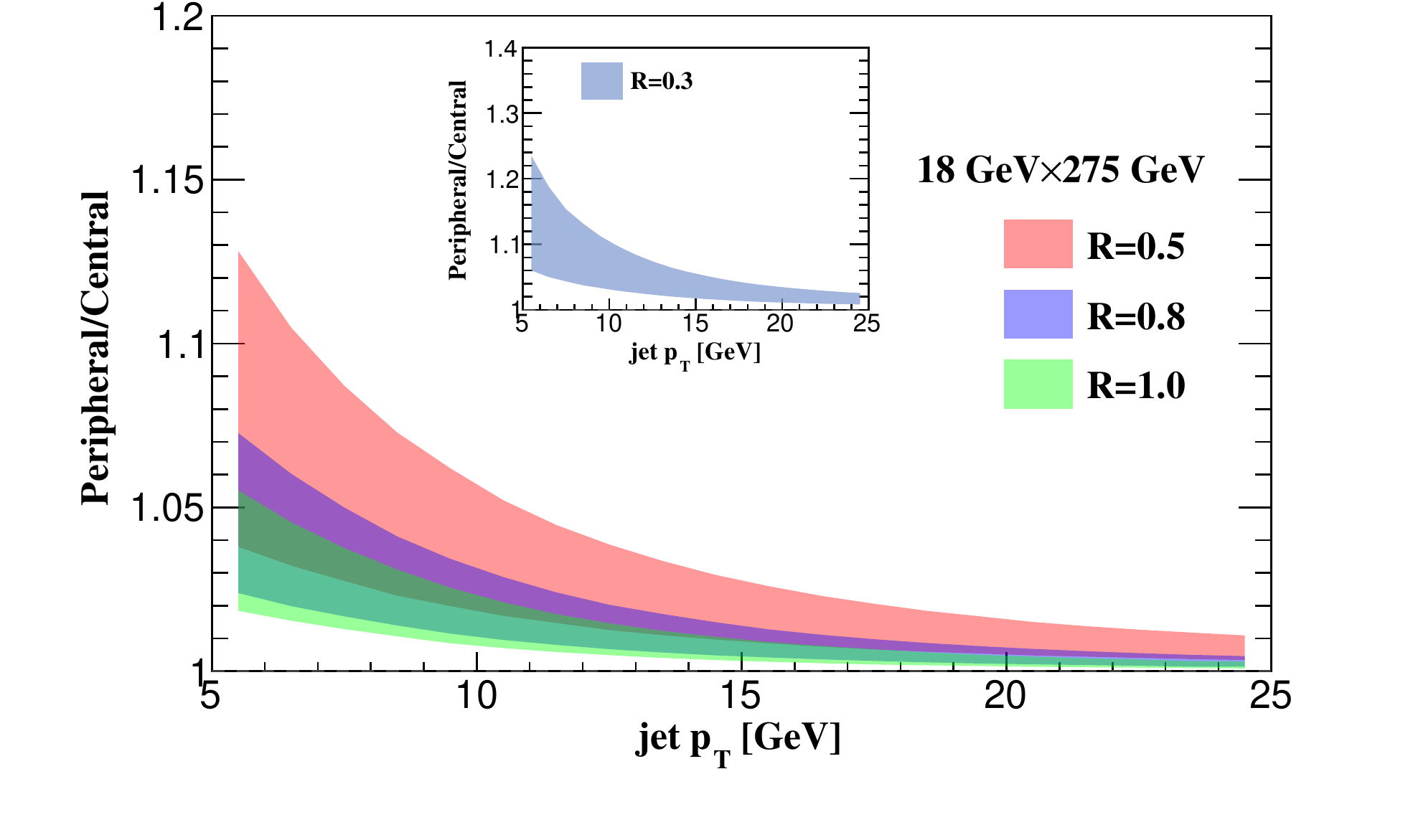}
    \vspace{-0.7cm}
    \caption{Ratio of per-nucleon jet  cross sections in peripheral and central collisions for the  jet  rapidity  interval is $2<\eta<4$  and  R=0.3 (inset), 0.5 (red), 0.8 (blue), and 1.0 (green). The upper  panel is for  10 $\times$ 100~GeV  $e$Pb collisions and the bottom one is for  18 $\times$ 275~GeV  $e$Pb collisions. }
    \label{fig:periToCent}
\end{figure}

Another method to directly investigate the interaction length dependence of final-state cold nuclear matter effects  using jet production is to compare the cross sections in peripheral and central collisions. Thus, we define the ratio as 
\begin{align} \label{eq:rp2c}
   \frac{\rm Peripheral}{\rm Central}(J) =
   \frac{ \frac{1}{\Delta_b T_A(b)}  \int_{\eta1}^{\eta2} \frac{d\sigma}{d\eta dp_T }\bigskip
    |_{e{\rm A},{\rm Peri.}}}{ \frac{1}{\Delta_b T_A(b)}  \int_{\eta1}^{\eta2} \frac{d\sigma}{d\eta dp_T}\bigskip
    |_{e{\rm A},{\rm Cent.}}},
\end{align}
where the initial-state effects  are reduced and most of the contribution  is from  final-state interactions. As discussed above, the medium induced energy loss is smaller and the per-nucleon cross section is larger for  peripheral collisions. Thus, the ratio defined in Eq.~(\ref{eq:rp2c}) is expected to be larger than one. 
Figure~\ref{fig:periToCent} displays our predictions for 10 GeV ($e$) $\times$ 100  GeV (Pb) and 18 GeV ($e$) $\times$ 275  GeV (Pb) collisions in the forward rapidity region for various jet radii. The $R=0.3$ case is shown in the insets since,  as expected, the ratio is much larger than in other cases and is very sensitive to the thickness of the nuclear matter in kinematic regions  where the jet $p_T$ distribution is steeper in particular when the collision energy is small. 
For  10 GeV ($e$) $\times$ 100  GeV (Pb) collisions the ratio can be around 1.1 for $R=0.5$ and $R=0.8$ in the small jet $p_T$  region.  It shows only about a few percent deviation from one for $R=1$.  The ratio in the large $p_T$ region is enhanced since  jets are produced close to the edges of phase space.    
For 18 GeV ($e$) $\times$ 275  GeV (Pb) collisions, the ratio  decreases with increasing jet $p_T$ and is smaller than 1.1 for most of the cases. 
The $R$ dependence indicates that the energy loss for larger radii is smaller which is consistent with Fig.~\ref{fig:ReAReA1}. In summary, the centrality class-dependent modification in matter is clearly observed in Fig.~\ref{fig:periToCent}.

 \begin{figure*}[t!]
 	\centering
 	\includegraphics[width=0.48\textwidth]{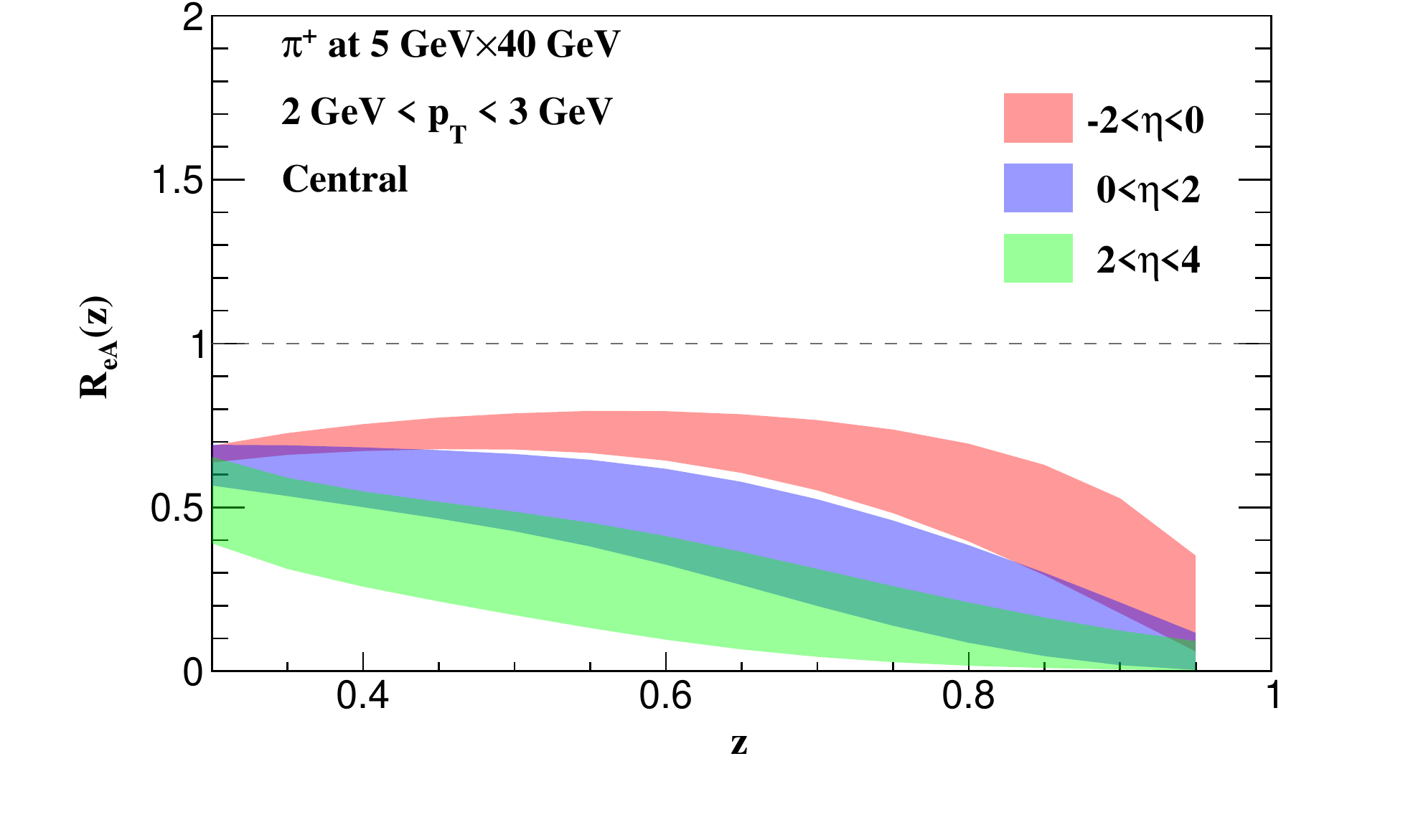}\,\,\,
 	\includegraphics[width=0.48\textwidth]{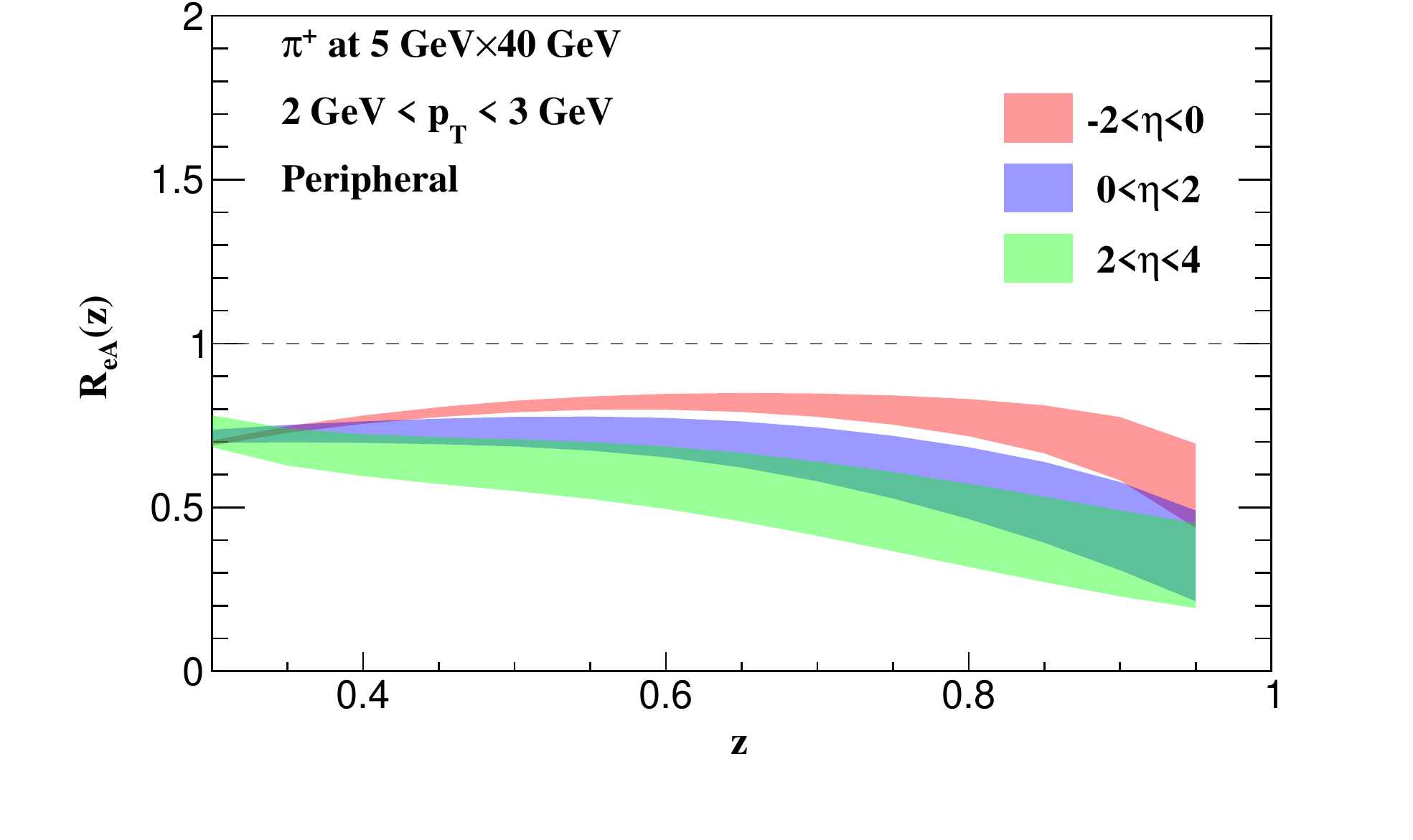}
 	\includegraphics[width=0.48\textwidth]{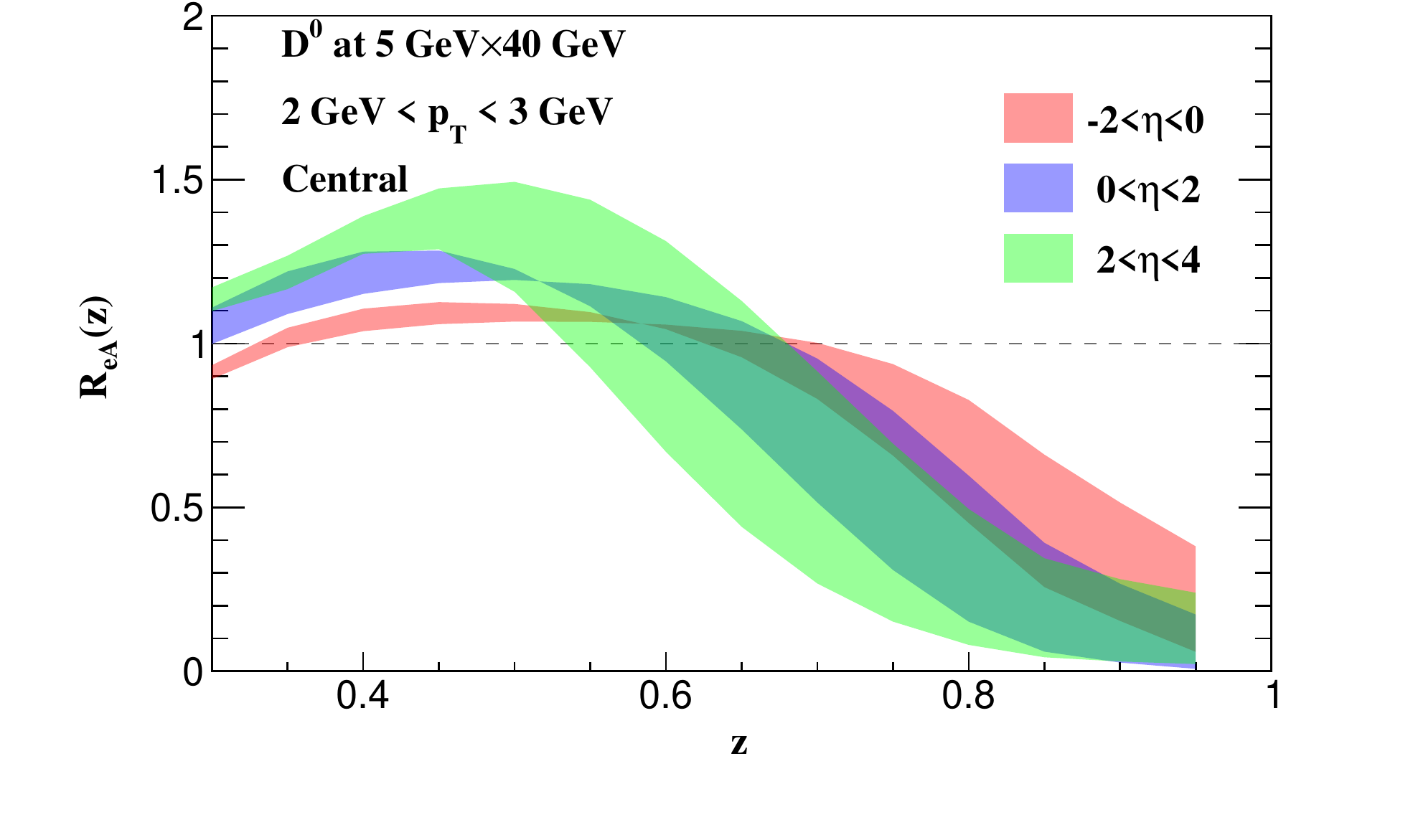}	\,\,\,
 	\includegraphics[width=0.48\textwidth]{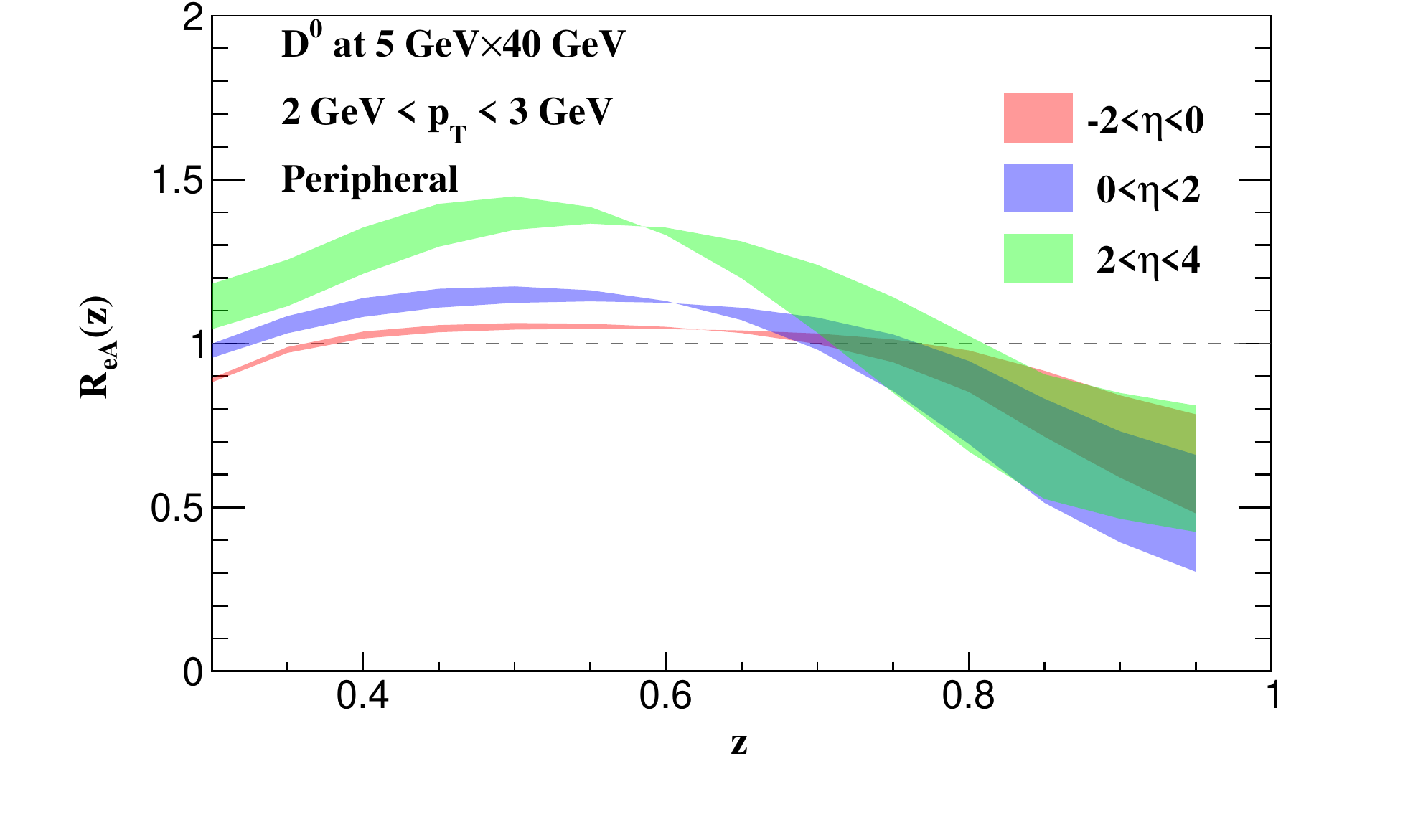}	
	\includegraphics[width=0.48\textwidth]{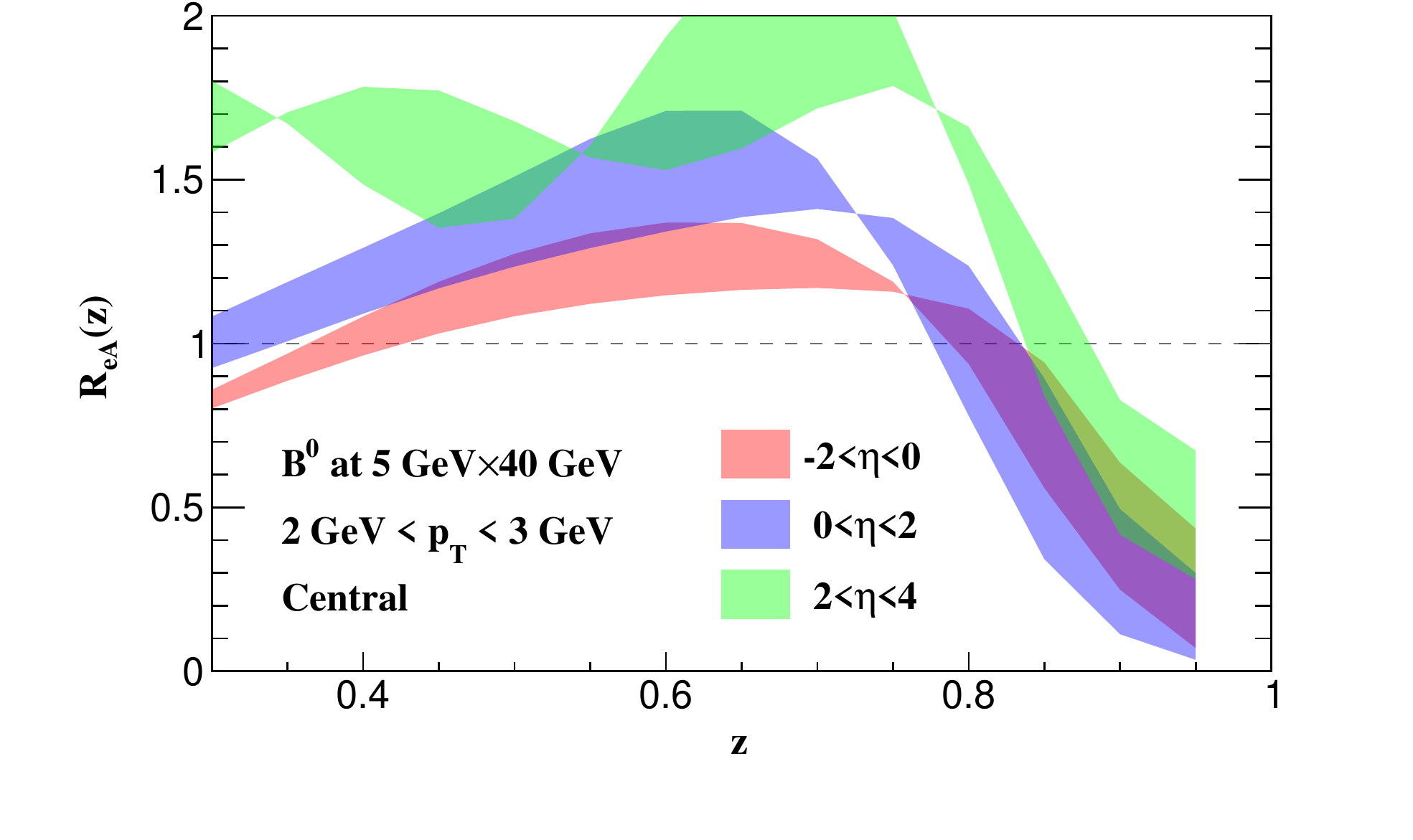}	\,\,\,
 	\includegraphics[width=0.48\textwidth]{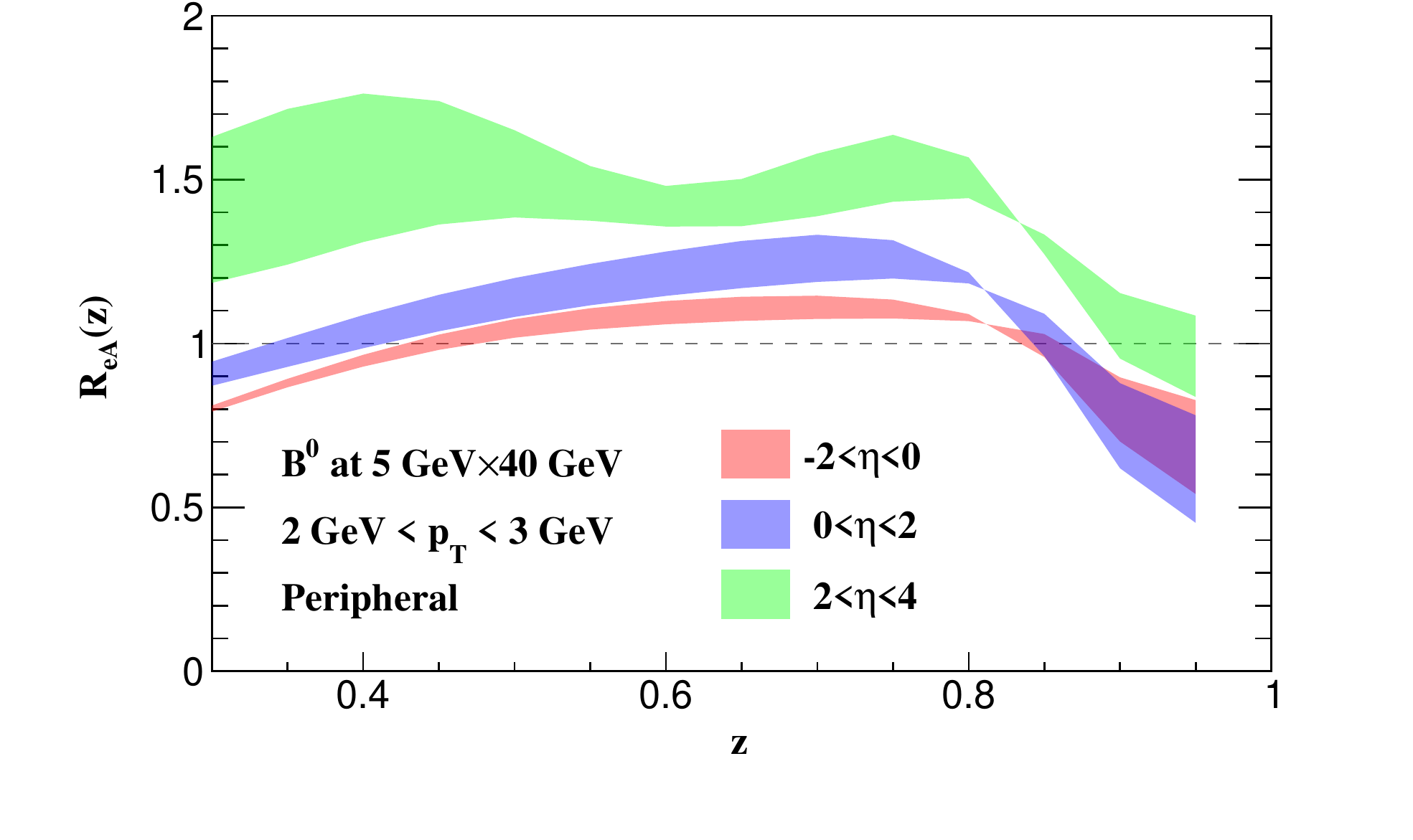}	
\caption{ In-medium corrections to $\pi^+$ (top panels), $D^0$ (middle panels) and $B^0$ (bottom panels) production as a function of $z$ at the EIC  in  5 GeV(e) $\times$ 40 GeV(A) collision. Red bands, blue bands, and  green bands correspond to  -2$<\eta<$0, 0$<\eta<$2 and 2$<\eta<$4, respectively. Results for central  collisions ($0-10\%$ centrality) are shown on the left, and results for peripheral  collisions ($80-100\%$ centrality) are shown on the right. }
 	\label{fig:zdisEIC_D0B0}
 \end{figure*}

Next, we discuss the cross-section modification for hadron production at the EIC, including $\pi^+$ and the  heavy $D_0$ and $B_0$ mesons. As shown in \cite{Li:2020zbk}, the following double ratio as a function of momentum fraction $z$ is a suitable observable for cold nuclear matter tomography at the EIC 
\begin{equation}
R_{eA}^h(z)=
\frac{
\frac{N^h(p_T,\eta,z)}{N^{\rm inc}(p_T,\eta)}\big|_{e\rm A}
}{
\frac{N^h(p_T,\eta,z)}{N^{\rm inc}(p_T,\eta)}\big|_{ep}
} \;.
\end{equation}
Here, we use the shorthand  notation  $N^h(p_T,\eta,z) \equiv  d \sigma^h/d\eta dp_T dz$ for the  distribution of hadrons versus the hadronization fraction $z$ and  $N^{\rm inc}(p_T,\eta) \equiv  d \sigma^J/d\eta dp_T $ for the  distribution of large radius jets.  In practice we integrate over suitably chosen rapidity and transverse momentum bins before taking the ratio. The idea behind normalizing by  $N^{\rm inc}(p_T,\eta)$ is to once again minimize initial-state effects and emphasize physics of final-state interactions in nuclear matter. For $e$A collisions we can further define per-nucleon cross sections by dividing out the $1/\Delta_b T_{\rm A}(b)$ geometric factor.

\begin{figure}[t!]
 	\centering
 	\includegraphics[width=0.48\textwidth]{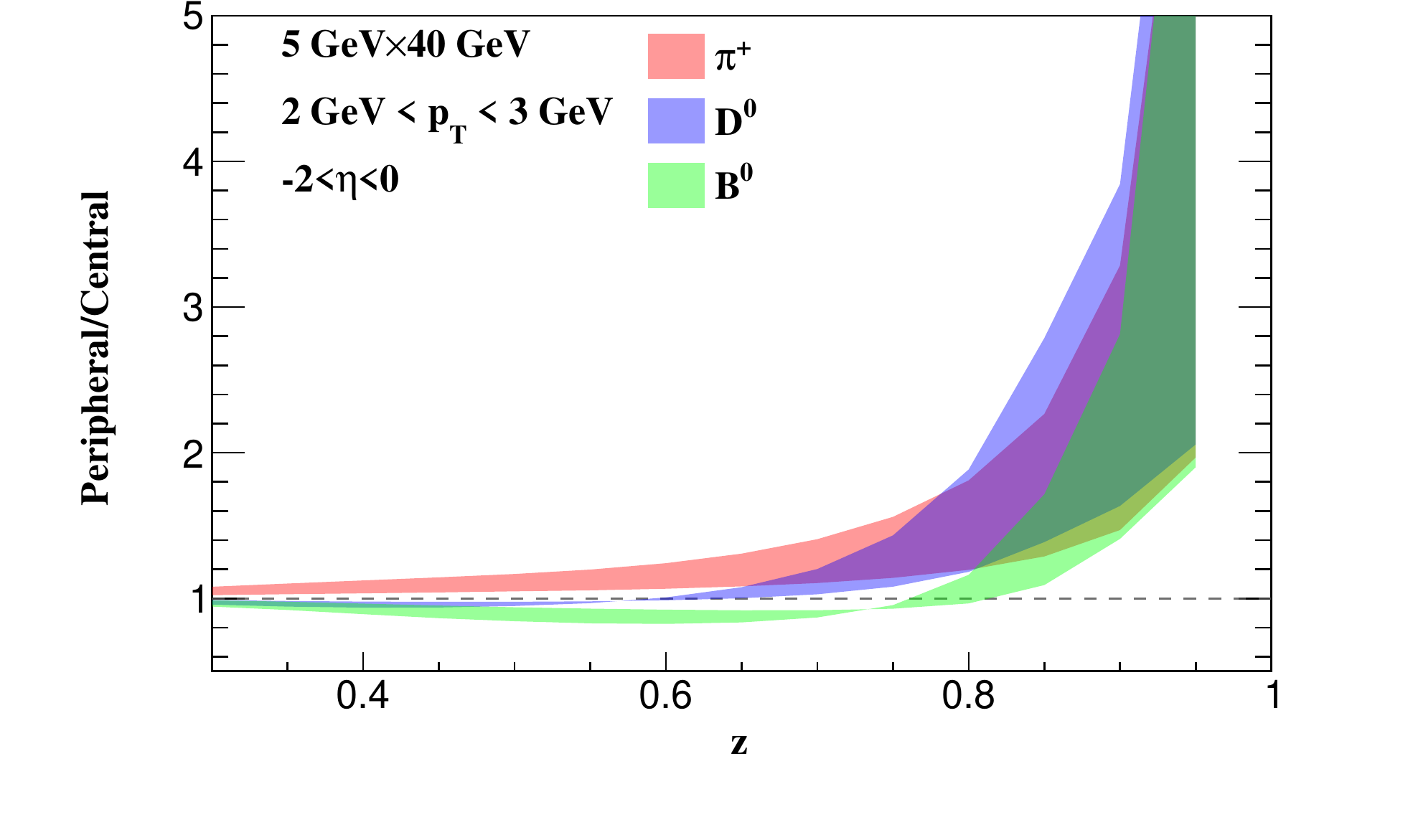}
 	\includegraphics[width=0.48\textwidth]{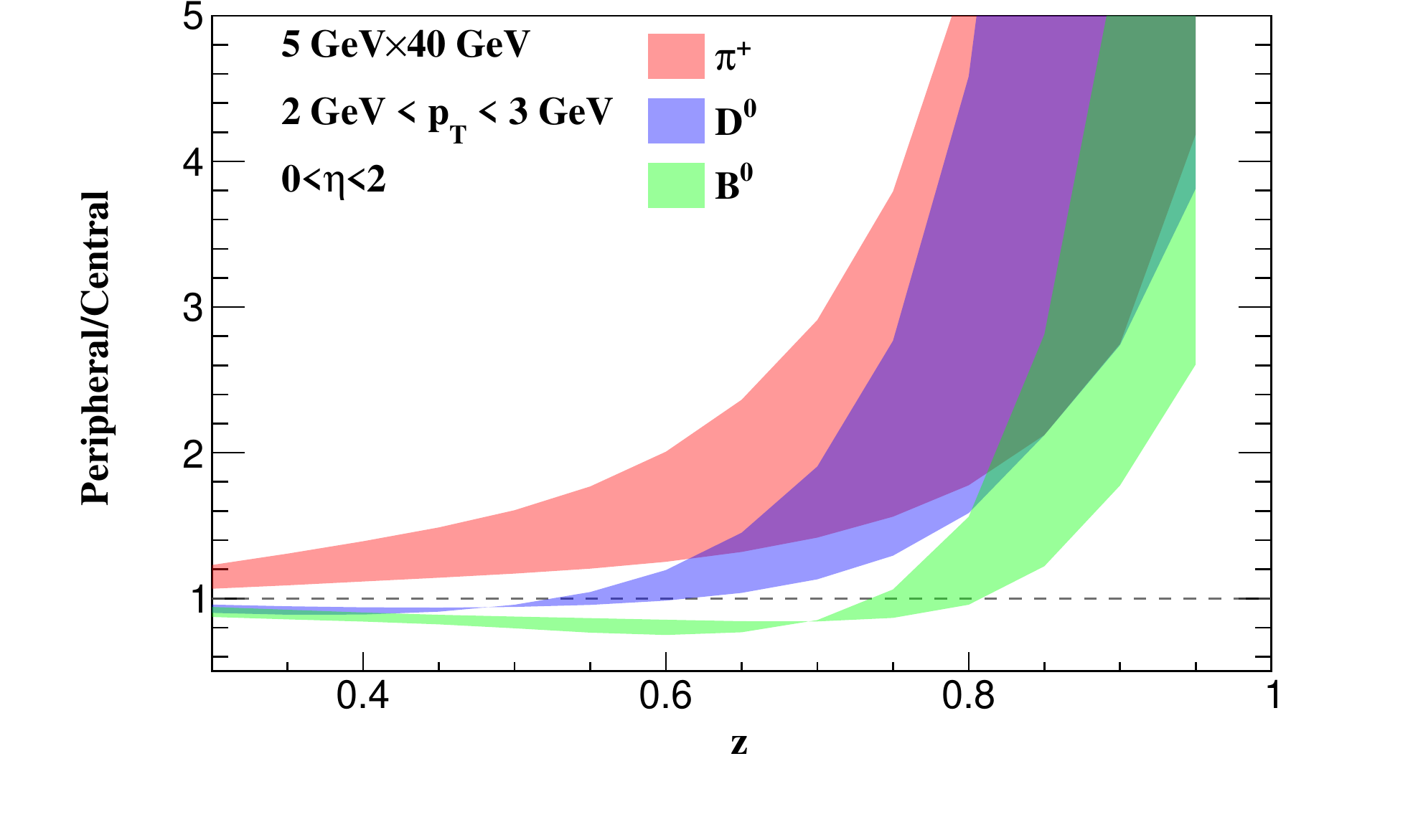}
 	\includegraphics[width=0.48\textwidth]{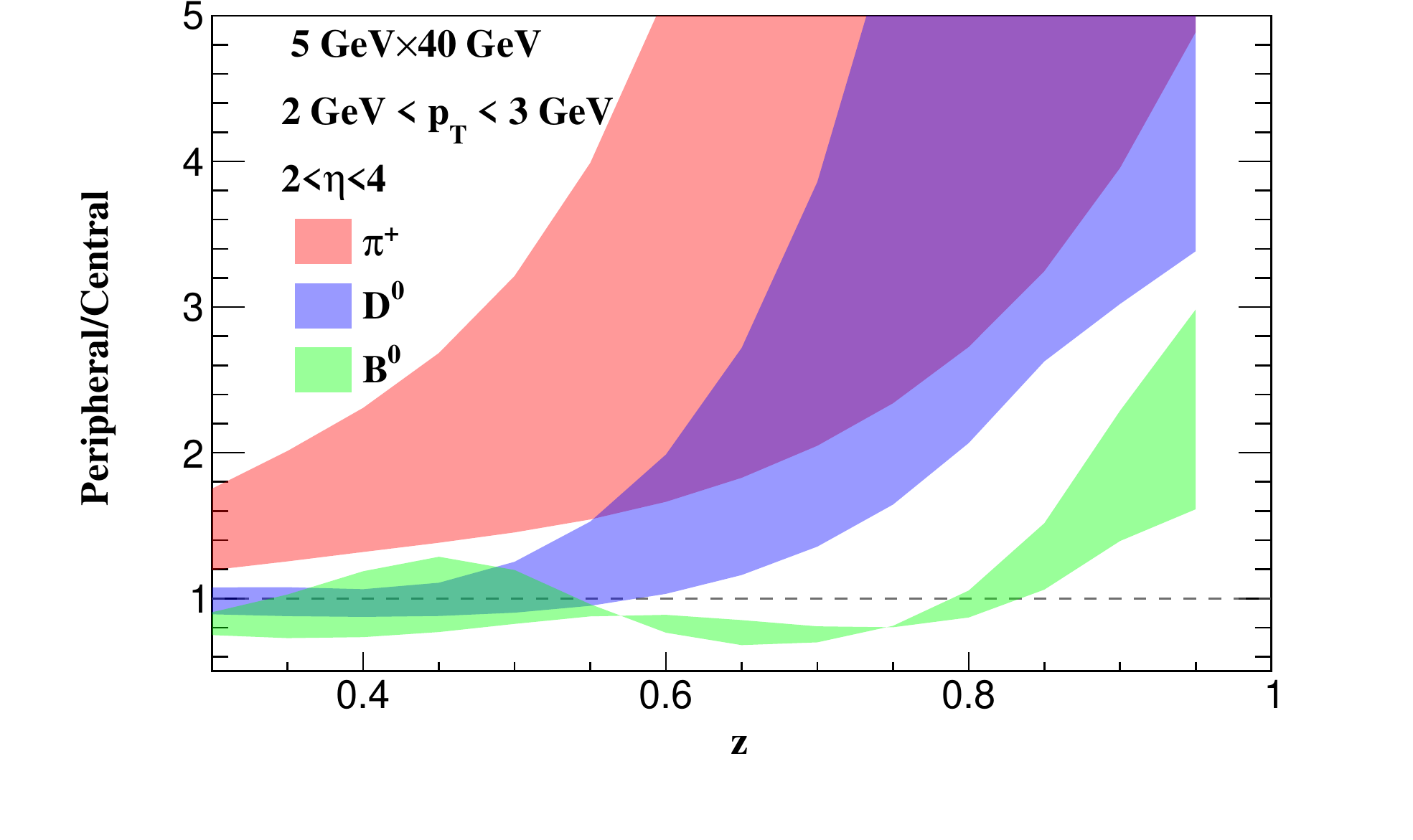}	
\caption{The ratio of $R_{eA}^h$ in peripheral  to central collisions. The electron and proton/nucleus beam energies, $p_T$ and $\eta$ ranges are the same as in  Fig.~\ref{fig:zdisEIC_D0B0}.  We show  $\pi^+$ (red), $D^0$ (blue) and $B^0$ (green). From top to bottom panels cover backward to forward rapidities.}
 	\label{fig:zdisEIC_cpr}
 \end{figure}

Our results for  $R_{e \rm Pb}^h(z)$ are shown in Fig.~\ref{fig:zdisEIC_cpr}.
 We consider electron-proton/nucleus collisions with energy 5 GeV (e) $\times$ 40 GeV (A), and the transverse momenta of final-state hadrons are fixed in the range 2 GeV to 3 GeV. Consequently,  the momentum fraction  $z$ distribution corresponds to the variation of $\nu$ constrained by the kinematics of the scattered electron in  experiment. The left column of panels is for the $0-10\%$ central events, and the right column is for $80-100\%$ peripheral events. For each hadron species, the red, blue and green bands correspond to  the predictions in rapidity regions $-2<\eta<0$, $0<\eta<2$ and $2<\eta<4$, respectively. Just as in the case of jet production, the bands reflect the variation of the nuclear matter transport parameter by a factor of two. Top to bottom rows show the differential  $\pi^+$, $D^0$ and $B^0$ modification. 
Because  lower energy partons receive larger medium corrections induced by the final-state interactions in the nucleus, the medium modification is more significant in the forward rapidity region $2 <\eta< 4$. In this region the energy of the final-state parton is lower in the nuclear rest frame in comparison, for example, to backward rapidity. It is instructive to observe that for light hadrons at large $z$ the differential cross section suppression can reach a factor of two even in peripheral collisions. In central events the energy loss effect can lead to more than an order of magnitude reduction.  For heavy flavor, just as in minimum bias reactions~\cite{Li:2020zbk},  $R_{e \rm Pb}^h(z)$ shows transition from suppression at large $z$ to enhancement at small $z$ because of the non-monotonic behavior of the heavy quark fragmentation function into heavy mesons~\cite{Braaten:1994bz,Cheung:1995ye}. In central reactions nuclear effects are noticeably larger.

To compare the  cross section modification in central and peripheral collisions for differential hadron distributions quantitatively, we define 
\begin{align} \label{eq:rp2c}
   \frac{\rm Peripheral}{\rm Central}(h) =
   \frac{  R_{eA}^h(z)  \bigskip
    |_{e{\rm A},{\rm Peri.}}}{ R_{eA}^h(z) \bigskip
    |_{e{\rm A},{\rm Cent.}}}
\end{align}
and note that the baseline $ep$ cross sections will drop out.  
As we expect, central collisions result in more significant medium corrections than peripheral ones, as shown in Fig.~\ref{fig:zdisEIC_D0B0}. The steep fragmentation distribution when $z \rightarrow 1$  enhances  the differences for light pions to  an order of magnitude. As we go forward in rapidity the enhancement in  ${\rm Peripheral}/{\rm Central}(h) $ extends to smaller $z$. For $D^0$ mesons this enhancement can also be very significant when $z\rightarrow 1$ but at intermediate fragmentation fractions the double ratio can dip below unity -- a consequence of the transition from suppression to enhancement in 
$R_{eA}^h(z)$. The qualitative behavior is is similar for $B^0$ mesons.

\section{Conclusions}  
\label{sec:concl}

We presented theoretical predictions for the nuclear modification of semi-inclusive hadron and jet production in $e$Pb collisions at the EIC  as a function of centrality.  We took advantage of recent simulations that were able to demonstrate robust correlation between centrality classes in $e$A and energy deposition in the zero-degree calorimeter, and to determine the mean interaction length seen by partons.  We constructed observables that minimize  initial-state nPDF effects and are sensitive to the inelastic final-state interactions of the struck parton in the nucleus. Future measurements of these observables at the EIC can provide essential information on the path length dependence of parton shower formation and hadronization in cold nuclear matter.

Our theoretical results indicate that the dependence of in-medium shower formation and energy loss on the transport properties and size of the nuclear medium can be easily identified and studied at the EIC. The exact sensitivity, however, depends on the choice of observables. We found that  for inclusive jets of small radius at moderate center-of-mass energies and at forward rapidities the per-nucleon cross sections variation between 0-10\% and 80-100\% collision can reach a factor of two. Because of the high integrated luminosity that EIC is expected to deliver~\cite{AbdulKhalek:2021gbh}, such peripheral-to-central differences will be easily measurable, but they are smaller than the differences in the  mean interaction length $\langle d \rangle$ seen by the jet. The reason for this is that even for $R = 0.3$ only a fraction of the medium-induced shower is redistributed outside of the jet cone.

Hadron measurements at forward rapidity can be performed at even lower center-of-mass energies. Our theoretical calculations showed that the per-nucleon differential particle distributions versus the fragmentation fraction $z_h$  depend much more significantly on centrality. For light pions at large $z_h$ the peripheral-to-central ratio  can reach a factor of 10, exceeding the ratio of effective interaction lengths for these centrality classes. Furthermore, the nuclear modification due to final-state interactions and its centrality variation are strong enough to be detected near mid rapidity and even at backward rapidity. The nuclear cross section modification also depends on the hadron flavor and has a predicted non-monotonic behavior for $D$- and $B$-mesons. We conclude by pointing out that in the future it will be important to explore the centrality dependence of other more differential jet observables such as jet substructure.  \\

\begin{acknowledgments}
The authors wish to thank R. Dupre for suggesting the calculation of centrality-dependent  hadron and  jet quenching at the EIC and P. Zurita for bringing early EMC measurements to our attention.  
We further acknowledge  helpful discussion with W. Chang and M. Baker on centrality determination in DIS and thank W. Chang for providing us with the effective interaction  lengths $\langle d \rangle$  in $e$A obtained with BeAGLE. 
I. Vitev is supported by the  LDRD program at LANL and by the U.S. Department of Energy under Contract No. 89233218CNA000001. 
H.T. Li is supported  by  the National Science Foundation of China under grant  No. 12275156. 
Z.L. Liu is funded by the European Union (ERC, grant agreement No. 101044599, JANUS). Views and opinions expressed are however those of the authors only and do not necessarily reflect those of the European Union or the European Research Council Executive Agency. Neither the European Union nor the granting authority can be held responsible for them.
 \end{acknowledgments}

\bibliography{DIScentral.bbl}

\begin{thebibliography}{67}
\expandafter\ifx\csname natexlab\endcsname\relax\def\natexlab#1{#1}\fi
\expandafter\ifx\csname bibnamefont\endcsname\relax
  \def\bibnamefont#1{#1}\fi
\expandafter\ifx\csname bibfnamefont\endcsname\relax
  \def\bibfnamefont#1{#1}\fi
\expandafter\ifx\csname citenamefont\endcsname\relax
  \def\citenamefont#1{#1}\fi
\expandafter\ifx\csname url\endcsname\relax
  \def\url#1{\texttt{#1}}\fi
\expandafter\ifx\csname urlprefix\endcsname\relax\def\urlprefix{URL }\fi
\providecommand{\bibinfo}[2]{#2}
\providecommand{\eprint}[2][]{\url{#2}}

\bibitem[{\citenamefont{Aubert et~al.}(1983)\citenamefont{Aubert, Bassompierre,
  Becks, Best, B{\"o}hm, {de Bouard}, Brasse, Broll, Brown, Carr
  et~al.}}]{AUBERT1983275}
\bibinfo{author}{\bibfnamefont{J.}~\bibnamefont{Aubert}},
  \bibinfo{author}{\bibfnamefont{G.}~\bibnamefont{Bassompierre}},
  \bibinfo{author}{\bibfnamefont{K.}~\bibnamefont{Becks}},
  \bibinfo{author}{\bibfnamefont{C.}~\bibnamefont{Best}},
  \bibinfo{author}{\bibfnamefont{E.}~\bibnamefont{B{\"o}hm}},
  \bibinfo{author}{\bibfnamefont{X.}~\bibnamefont{{de Bouard}}},
  \bibinfo{author}{\bibfnamefont{F.}~\bibnamefont{Brasse}},
  \bibinfo{author}{\bibfnamefont{C.}~\bibnamefont{Broll}},
  \bibinfo{author}{\bibfnamefont{S.}~\bibnamefont{Brown}},
  \bibinfo{author}{\bibfnamefont{J.}~\bibnamefont{Carr}}, \bibnamefont{et~al.},
  \bibinfo{journal}{Physics Letters B} \textbf{\bibinfo{volume}{123}},
  \bibinfo{pages}{275} (\bibinfo{year}{1983}).

\bibitem[{Pro(2020)}]{Proceedings:2020eah}
\emph{\bibinfo{title}{{Proceedings, Probing Nucleons and Nuclei in High Energy
  Collisions: Dedicated to the Physics of the Electron Ion Collider}: {Seattle
  (WA), United States, October 1 - November 16, 2018}}}
  (\bibinfo{publisher}{WSP}, \bibinfo{year}{2020}), \eprint{2002.12333}.

\bibitem[{\citenamefont{Abdul~Khalek et~al.}(2022)}]{AbdulKhalek:2021gbh}
\bibinfo{author}{\bibfnamefont{R.}~\bibnamefont{Abdul~Khalek}}
  \bibnamefont{et~al.}, \bibinfo{journal}{Nucl. Phys. A}
  \textbf{\bibinfo{volume}{1026}}, \bibinfo{pages}{122447}
  (\bibinfo{year}{2022}), \eprint{2103.05419}.

\bibitem[{\citenamefont{Albacete et~al.}(2013)}]{Albacete:2013ei}
\bibinfo{author}{\bibfnamefont{J.~L.} \bibnamefont{Albacete}}
  \bibnamefont{et~al.}, \bibinfo{journal}{Int. J. Mod. Phys. E}
  \textbf{\bibinfo{volume}{22}}, \bibinfo{pages}{1330007}
  (\bibinfo{year}{2013}), \eprint{1301.3395}.

\bibitem[{\citenamefont{Albacete et~al.}(2018)}]{Albacete:2017qng}
\bibinfo{author}{\bibfnamefont{J.~L.} \bibnamefont{Albacete}}
  \bibnamefont{et~al.}, \bibinfo{journal}{Nucl. Phys. A}
  \textbf{\bibinfo{volume}{972}}, \bibinfo{pages}{18} (\bibinfo{year}{2018}),
  \eprint{1707.09973}.

\bibitem[{\citenamefont{Kovarik et~al.}(2016)}]{Kovarik:2015cma}
\bibinfo{author}{\bibfnamefont{K.}~\bibnamefont{Kovarik}} \bibnamefont{et~al.},
  \bibinfo{journal}{Phys. Rev.} \textbf{\bibinfo{volume}{D93}},
  \bibinfo{pages}{085037} (\bibinfo{year}{2016}), \eprint{1509.00792}.

\bibitem[{\citenamefont{Eskola et~al.}(2017)\citenamefont{Eskola, Paakkinen,
  Paukkunen, and Salgado}}]{Eskola:2016oht}
\bibinfo{author}{\bibfnamefont{K.~J.} \bibnamefont{Eskola}},
  \bibinfo{author}{\bibfnamefont{P.}~\bibnamefont{Paakkinen}},
  \bibinfo{author}{\bibfnamefont{H.}~\bibnamefont{Paukkunen}},
  \bibnamefont{and} \bibinfo{author}{\bibfnamefont{C.~A.}
  \bibnamefont{Salgado}}, \bibinfo{journal}{Eur. Phys. J.}
  \textbf{\bibinfo{volume}{C77}}, \bibinfo{pages}{163} (\bibinfo{year}{2017}),
  \eprint{1612.05741}.

\bibitem[{\citenamefont{Abdul~Khalek et~al.}(2019)\citenamefont{Abdul~Khalek,
  Ethier, and Rojo}}]{AbdulKhalek:2019mzd}
\bibinfo{author}{\bibfnamefont{R.}~\bibnamefont{Abdul~Khalek}},
  \bibinfo{author}{\bibfnamefont{J.~J.} \bibnamefont{Ethier}},
  \bibnamefont{and} \bibinfo{author}{\bibfnamefont{J.}~\bibnamefont{Rojo}}
  (\bibinfo{collaboration}{NNPDF}), \bibinfo{journal}{Eur. Phys. J. C}
  \textbf{\bibinfo{volume}{79}}, \bibinfo{pages}{471} (\bibinfo{year}{2019}),
  \eprint{1904.00018}.

\bibitem[{\citenamefont{Balitsky and Lipatov}(1978)}]{Balitsky:1978ic}
\bibinfo{author}{\bibfnamefont{I.~I.} \bibnamefont{Balitsky}} \bibnamefont{and}
  \bibinfo{author}{\bibfnamefont{L.~N.} \bibnamefont{Lipatov}},
  \bibinfo{journal}{Sov. J. Nucl. Phys.} \textbf{\bibinfo{volume}{28}},
  \bibinfo{pages}{822} (\bibinfo{year}{1978}), \bibinfo{note}{[Yad.
  Fiz.28,1597(1978)]}.

\bibitem[{\citenamefont{McLerran and Venugopalan}(1994)}]{McLerran:1993ka}
\bibinfo{author}{\bibfnamefont{L.~D.} \bibnamefont{McLerran}} \bibnamefont{and}
  \bibinfo{author}{\bibfnamefont{R.}~\bibnamefont{Venugopalan}},
  \bibinfo{journal}{Phys. Rev.} \textbf{\bibinfo{volume}{D49}},
  \bibinfo{pages}{3352} (\bibinfo{year}{1994}), \eprint{hep-ph/9311205}.

\bibitem[{\citenamefont{Jalilian-Marian
  et~al.}(1998)\citenamefont{Jalilian-Marian, Kovner, Leonidov, and
  Weigert}}]{JalilianMarian:1997gr}
\bibinfo{author}{\bibfnamefont{J.}~\bibnamefont{Jalilian-Marian}},
  \bibinfo{author}{\bibfnamefont{A.}~\bibnamefont{Kovner}},
  \bibinfo{author}{\bibfnamefont{A.}~\bibnamefont{Leonidov}}, \bibnamefont{and}
  \bibinfo{author}{\bibfnamefont{H.}~\bibnamefont{Weigert}},
  \bibinfo{journal}{Phys. Rev.} \textbf{\bibinfo{volume}{D59}},
  \bibinfo{pages}{014014} (\bibinfo{year}{1998}), \eprint{hep-ph/9706377}.

\bibitem[{\citenamefont{Kovchegov}(1999)}]{Kovchegov:1999yj}
\bibinfo{author}{\bibfnamefont{Y.~V.} \bibnamefont{Kovchegov}},
  \bibinfo{journal}{Phys. Rev.} \textbf{\bibinfo{volume}{D60}},
  \bibinfo{pages}{034008} (\bibinfo{year}{1999}), \eprint{hep-ph/9901281}.

\bibitem[{\citenamefont{Wang and Wang}(2002)}]{Wang:2002ri}
\bibinfo{author}{\bibfnamefont{E.}~\bibnamefont{Wang}} \bibnamefont{and}
  \bibinfo{author}{\bibfnamefont{X.-N.} \bibnamefont{Wang}},
  \bibinfo{journal}{Phys. Rev. Lett.} \textbf{\bibinfo{volume}{89}},
  \bibinfo{pages}{162301} (\bibinfo{year}{2002}), \eprint{hep-ph/0202105}.

\bibitem[{\citenamefont{Qiu and Vitev}(2004)}]{Qiu:2004qk}
\bibinfo{author}{\bibfnamefont{J.-W.} \bibnamefont{Qiu}} \bibnamefont{and}
  \bibinfo{author}{\bibfnamefont{I.}~\bibnamefont{Vitev}},
  \bibinfo{journal}{Phys. Lett. B} \textbf{\bibinfo{volume}{587}},
  \bibinfo{pages}{52} (\bibinfo{year}{2004}), \eprint{hep-ph/0401062}.

\bibitem[{\citenamefont{Vitev}(2007)}]{Vitev:2007ve}
\bibinfo{author}{\bibfnamefont{I.}~\bibnamefont{Vitev}},
  \bibinfo{journal}{Phys. Rev. C} \textbf{\bibinfo{volume}{75}},
  \bibinfo{pages}{064906} (\bibinfo{year}{2007}), \eprint{hep-ph/0703002}.

\bibitem[{\citenamefont{Arratia et~al.}(2020)\citenamefont{Arratia, Song,
  Ringer, and Jacak}}]{Arratia:2019vju}
\bibinfo{author}{\bibfnamefont{M.}~\bibnamefont{Arratia}},
  \bibinfo{author}{\bibfnamefont{Y.}~\bibnamefont{Song}},
  \bibinfo{author}{\bibfnamefont{F.}~\bibnamefont{Ringer}}, \bibnamefont{and}
  \bibinfo{author}{\bibfnamefont{B.~V.} \bibnamefont{Jacak}},
  \bibinfo{journal}{Phys. Rev. C} \textbf{\bibinfo{volume}{101}},
  \bibinfo{pages}{065204} (\bibinfo{year}{2020}), \eprint{1912.05931}.

\bibitem[{\citenamefont{Gavin and Milana}(1992)}]{Gavin:1991qk}
\bibinfo{author}{\bibfnamefont{S.}~\bibnamefont{Gavin}} \bibnamefont{and}
  \bibinfo{author}{\bibfnamefont{J.}~\bibnamefont{Milana}},
  \bibinfo{journal}{Phys. Rev. Lett.} \textbf{\bibinfo{volume}{68}},
  \bibinfo{pages}{1834} (\bibinfo{year}{1992}).

\bibitem[{\citenamefont{Arleo}(2002)}]{Arleo:2002ph}
\bibinfo{author}{\bibfnamefont{F.}~\bibnamefont{Arleo}},
  \bibinfo{journal}{Phys. Lett. B} \textbf{\bibinfo{volume}{532}},
  \bibinfo{pages}{231} (\bibinfo{year}{2002}), \eprint{hep-ph/0201066}.

\bibitem[{\citenamefont{Neufeld et~al.}(2011)\citenamefont{Neufeld, Vitev, and
  Zhang}}]{Neufeld:2010dz}
\bibinfo{author}{\bibfnamefont{R.~B.} \bibnamefont{Neufeld}},
  \bibinfo{author}{\bibfnamefont{I.}~\bibnamefont{Vitev}}, \bibnamefont{and}
  \bibinfo{author}{\bibfnamefont{B.-W.} \bibnamefont{Zhang}},
  \bibinfo{journal}{Phys. Lett. B} \textbf{\bibinfo{volume}{704}},
  \bibinfo{pages}{590} (\bibinfo{year}{2011}), \eprint{1010.3708}.

\bibitem[{\citenamefont{Xing et~al.}(2012)\citenamefont{Xing, Guo, Wang, and
  Wang}}]{Xing:2011fb}
\bibinfo{author}{\bibfnamefont{H.}~\bibnamefont{Xing}},
  \bibinfo{author}{\bibfnamefont{Y.}~\bibnamefont{Guo}},
  \bibinfo{author}{\bibfnamefont{E.}~\bibnamefont{Wang}}, \bibnamefont{and}
  \bibinfo{author}{\bibfnamefont{X.-N.} \bibnamefont{Wang}},
  \bibinfo{journal}{Nucl. Phys. A} \textbf{\bibinfo{volume}{879}},
  \bibinfo{pages}{77} (\bibinfo{year}{2012}), \eprint{1110.1903}.

\bibitem[{\citenamefont{Arleo et~al.}(2019)\citenamefont{Arleo, Na\"\i{}m, and
  Platchkov}}]{Arleo:2018zjw}
\bibinfo{author}{\bibfnamefont{F.}~\bibnamefont{Arleo}},
  \bibinfo{author}{\bibfnamefont{C.-J.} \bibnamefont{Na\"\i{}m}},
  \bibnamefont{and}
  \bibinfo{author}{\bibfnamefont{S.}~\bibnamefont{Platchkov}},
  \bibinfo{journal}{JHEP} \textbf{\bibinfo{volume}{01}}, \bibinfo{pages}{129}
  (\bibinfo{year}{2019}), \eprint{1810.05120}.

\bibitem[{\citenamefont{Kang et~al.}(2015)\citenamefont{Kang, Vitev, and
  Xing}}]{Kang:2015mta}
\bibinfo{author}{\bibfnamefont{Z.-B.} \bibnamefont{Kang}},
  \bibinfo{author}{\bibfnamefont{I.}~\bibnamefont{Vitev}}, \bibnamefont{and}
  \bibinfo{author}{\bibfnamefont{H.}~\bibnamefont{Xing}},
  \bibinfo{journal}{Phys. Rev. C} \textbf{\bibinfo{volume}{92}},
  \bibinfo{pages}{054911} (\bibinfo{year}{2015}), \eprint{1507.05987}.

\bibitem[{\citenamefont{Johnson et~al.}(2001)}]{FNALE772:2000fmo}
\bibinfo{author}{\bibfnamefont{M.~B.} \bibnamefont{Johnson}}
  \bibnamefont{et~al.} (\bibinfo{collaboration}{FNAL E772}),
  \bibinfo{journal}{Phys. Rev. Lett.} \textbf{\bibinfo{volume}{86}},
  \bibinfo{pages}{4483} (\bibinfo{year}{2001}), \eprint{hep-ex/0010051}.

\bibitem[{\citenamefont{Leitch et~al.}(2000)\citenamefont{Leitch, Lee, Beddo,
  Brown, Carey, Chang, Cooper, Gagliardi, Garvey, Geesaman
  et~al.}}]{PhysRevLett.84.3256}
\bibinfo{author}{\bibfnamefont{M.~J.} \bibnamefont{Leitch}},
  \bibinfo{author}{\bibfnamefont{W.~M.} \bibnamefont{Lee}},
  \bibinfo{author}{\bibfnamefont{M.~E.} \bibnamefont{Beddo}},
  \bibinfo{author}{\bibfnamefont{C.~N.} \bibnamefont{Brown}},
  \bibinfo{author}{\bibfnamefont{T.~A.} \bibnamefont{Carey}},
  \bibinfo{author}{\bibfnamefont{T.~H.} \bibnamefont{Chang}},
  \bibinfo{author}{\bibfnamefont{W.~E.} \bibnamefont{Cooper}},
  \bibinfo{author}{\bibfnamefont{C.~A.} \bibnamefont{Gagliardi}},
  \bibinfo{author}{\bibfnamefont{G.~T.} \bibnamefont{Garvey}},
  \bibinfo{author}{\bibfnamefont{D.~F.} \bibnamefont{Geesaman}},
  \bibnamefont{et~al.} (\bibinfo{collaboration}{FNAL E866/NuSea
  Collaboration}), \bibinfo{journal}{Phys. Rev. Lett.}
  \textbf{\bibinfo{volume}{84}}, \bibinfo{pages}{3256} (\bibinfo{year}{2000}).

\bibitem[{\citenamefont{Aad et~al.}(2015)}]{ATLAS:2014cpa}
\bibinfo{author}{\bibfnamefont{G.}~\bibnamefont{Aad}} \bibnamefont{et~al.}
  (\bibinfo{collaboration}{ATLAS}), \bibinfo{journal}{Phys. Lett. B}
  \textbf{\bibinfo{volume}{748}}, \bibinfo{pages}{392} (\bibinfo{year}{2015}),
  \eprint{1412.4092}.

\bibitem[{\citenamefont{Adare et~al.}(2016)}]{PHENIX:2015fgy}
\bibinfo{author}{\bibfnamefont{A.}~\bibnamefont{Adare}} \bibnamefont{et~al.}
  (\bibinfo{collaboration}{PHENIX}), \bibinfo{journal}{Phys. Rev. Lett.}
  \textbf{\bibinfo{volume}{116}}, \bibinfo{pages}{122301}
  (\bibinfo{year}{2016}), \eprint{1509.04657}.

\bibitem[{\citenamefont{Arleo}(2003)}]{Arleo:2003jz}
\bibinfo{author}{\bibfnamefont{F.}~\bibnamefont{Arleo}}, \bibinfo{journal}{Eur.
  Phys. J. C} \textbf{\bibinfo{volume}{30}}, \bibinfo{pages}{213}
  (\bibinfo{year}{2003}), \eprint{hep-ph/0306235}.

\bibitem[{\citenamefont{Chang et~al.}(2014)\citenamefont{Chang, Deng, and
  Wang}}]{Chang:2014fba}
\bibinfo{author}{\bibfnamefont{N.-B.} \bibnamefont{Chang}},
  \bibinfo{author}{\bibfnamefont{W.-T.} \bibnamefont{Deng}}, \bibnamefont{and}
  \bibinfo{author}{\bibfnamefont{X.-N.} \bibnamefont{Wang}},
  \bibinfo{journal}{Phys. Rev.} \textbf{\bibinfo{volume}{C89}},
  \bibinfo{pages}{034911} (\bibinfo{year}{2014}), \eprint{1401.5109}.

\bibitem[{\citenamefont{Li et~al.}(2020)\citenamefont{Li, Liu, and
  Vitev}}]{Li:2020zbk}
\bibinfo{author}{\bibfnamefont{H.~T.} \bibnamefont{Li}},
  \bibinfo{author}{\bibfnamefont{Z.~L.} \bibnamefont{Liu}}, \bibnamefont{and}
  \bibinfo{author}{\bibfnamefont{I.}~\bibnamefont{Vitev}}
  (\bibinfo{year}{2020}), \eprint{2007.10994}.

\bibitem[{\citenamefont{Ke and Vitev}(2023)}]{Ke:2023ixa}
\bibinfo{author}{\bibfnamefont{W.}~\bibnamefont{Ke}} \bibnamefont{and}
  \bibinfo{author}{\bibfnamefont{I.}~\bibnamefont{Vitev}}
  (\bibinfo{year}{2023}), \eprint{2301.11940}.

\bibitem[{\citenamefont{Airapetian et~al.}(2001)}]{Airapetian:2000ks}
\bibinfo{author}{\bibfnamefont{A.}~\bibnamefont{Airapetian}}
  \bibnamefont{et~al.} (\bibinfo{collaboration}{HERMES}),
  \bibinfo{journal}{Eur. Phys. J. C} \textbf{\bibinfo{volume}{20}},
  \bibinfo{pages}{479} (\bibinfo{year}{2001}), \eprint{hep-ex/0012049}.

\bibitem[{\citenamefont{Airapetian et~al.}(2003)}]{Airapetian:2003mi}
\bibinfo{author}{\bibfnamefont{A.}~\bibnamefont{Airapetian}}
  \bibnamefont{et~al.} (\bibinfo{collaboration}{HERMES}),
  \bibinfo{journal}{Phys. Lett. B} \textbf{\bibinfo{volume}{577}},
  \bibinfo{pages}{37} (\bibinfo{year}{2003}), \eprint{hep-ex/0307023}.

\bibitem[{\citenamefont{Airapetian et~al.}(2007)}]{Airapetian:2007vu}
\bibinfo{author}{\bibfnamefont{A.}~\bibnamefont{Airapetian}}
  \bibnamefont{et~al.} (\bibinfo{collaboration}{HERMES}),
  \bibinfo{journal}{Nucl. Phys. B} \textbf{\bibinfo{volume}{780}},
  \bibinfo{pages}{1} (\bibinfo{year}{2007}), \eprint{0704.3270}.

\bibitem[{\citenamefont{Ashman et~al.}(1991)\citenamefont{Ashman, Badelek,
  Baum, Beaufays, Bee, Benchouk, Bird, Brown, Caputo, Cheung et~al.}}]{EMC1}
\bibinfo{author}{\bibfnamefont{J.}~\bibnamefont{Ashman}},
  \bibinfo{author}{\bibfnamefont{B.}~\bibnamefont{Badelek}},
  \bibinfo{author}{\bibfnamefont{G.}~\bibnamefont{Baum}},
  \bibinfo{author}{\bibfnamefont{J.}~\bibnamefont{Beaufays}},
  \bibinfo{author}{\bibfnamefont{C.~P.} \bibnamefont{Bee}},
  \bibinfo{author}{\bibfnamefont{C.}~\bibnamefont{Benchouk}},
  \bibinfo{author}{\bibfnamefont{I.~G.} \bibnamefont{Bird}},
  \bibinfo{author}{\bibfnamefont{S.~C.} \bibnamefont{Brown}},
  \bibinfo{author}{\bibfnamefont{M.~C.} \bibnamefont{Caputo}},
  \bibinfo{author}{\bibfnamefont{H.~W.~K.} \bibnamefont{Cheung}},
  \bibnamefont{et~al.}, \bibinfo{journal}{Zeitschrift f{\"u}r Physik C
  Particles and Fields} \textbf{\bibinfo{volume}{52}}, \bibinfo{pages}{1}
  (\bibinfo{year}{1991}).

\bibitem[{\citenamefont{Arvidson et~al.}(1984)\citenamefont{Arvidson, Aubert,
  Bassompierre, Becks, Benchouk, Best, Böhm, {de Bouard}, Brasse, Broll
  et~al.}}]{ARVIDSON1984381}
\bibinfo{author}{\bibfnamefont{A.}~\bibnamefont{Arvidson}},
  \bibinfo{author}{\bibfnamefont{J.}~\bibnamefont{Aubert}},
  \bibinfo{author}{\bibfnamefont{G.}~\bibnamefont{Bassompierre}},
  \bibinfo{author}{\bibfnamefont{K.}~\bibnamefont{Becks}},
  \bibinfo{author}{\bibfnamefont{C.}~\bibnamefont{Benchouk}},
  \bibinfo{author}{\bibfnamefont{C.}~\bibnamefont{Best}},
  \bibinfo{author}{\bibfnamefont{E.}~\bibnamefont{Böhm}},
  \bibinfo{author}{\bibfnamefont{X.}~\bibnamefont{{de Bouard}}},
  \bibinfo{author}{\bibfnamefont{F.}~\bibnamefont{Brasse}},
  \bibinfo{author}{\bibfnamefont{C.}~\bibnamefont{Broll}},
  \bibnamefont{et~al.}, \bibinfo{journal}{Nuclear Physics B}
  \textbf{\bibinfo{volume}{246}}, \bibinfo{pages}{381} (\bibinfo{year}{1984}),
  ISSN \bibinfo{issn}{0550-3213}.

\bibitem[{\citenamefont{Accardi et~al.}(2003)\citenamefont{Accardi, Muccifora,
  and Pirner}}]{Accardi:2002tv}
\bibinfo{author}{\bibfnamefont{A.}~\bibnamefont{Accardi}},
  \bibinfo{author}{\bibfnamefont{V.}~\bibnamefont{Muccifora}},
  \bibnamefont{and} \bibinfo{author}{\bibfnamefont{H.-J.}
  \bibnamefont{Pirner}}, \bibinfo{journal}{Nucl. Phys. A}
  \textbf{\bibinfo{volume}{720}}, \bibinfo{pages}{131} (\bibinfo{year}{2003}),
  \eprint{nucl-th/0211011}.

\bibitem[{\citenamefont{Kopeliovich et~al.}(2004)\citenamefont{Kopeliovich,
  Nemchik, Predazzi, and Hayashigaki}}]{Kopeliovich:2003py}
\bibinfo{author}{\bibfnamefont{B.}~\bibnamefont{Kopeliovich}},
  \bibinfo{author}{\bibfnamefont{J.}~\bibnamefont{Nemchik}},
  \bibinfo{author}{\bibfnamefont{E.}~\bibnamefont{Predazzi}}, \bibnamefont{and}
  \bibinfo{author}{\bibfnamefont{A.}~\bibnamefont{Hayashigaki}},
  \bibinfo{journal}{Nucl. Phys. A} \textbf{\bibinfo{volume}{740}},
  \bibinfo{pages}{211} (\bibinfo{year}{2004}), \eprint{hep-ph/0311220}.

\bibitem[{\citenamefont{Sassot et~al.}(2010)\citenamefont{Sassot, Stratmann,
  and Zurita}}]{Sassot:2009sh}
\bibinfo{author}{\bibfnamefont{R.}~\bibnamefont{Sassot}},
  \bibinfo{author}{\bibfnamefont{M.}~\bibnamefont{Stratmann}},
  \bibnamefont{and} \bibinfo{author}{\bibfnamefont{P.}~\bibnamefont{Zurita}},
  \bibinfo{journal}{Phys. Rev. D} \textbf{\bibinfo{volume}{81}},
  \bibinfo{pages}{054001} (\bibinfo{year}{2010}), \eprint{0912.1311}.

\bibitem[{\citenamefont{Zurita}(2021)}]{Zurita:2021kli}
\bibinfo{author}{\bibfnamefont{P.}~\bibnamefont{Zurita}}
  (\bibinfo{year}{2021}), \eprint{2101.01088}.

\bibitem[{\citenamefont{Accardi et~al.}(2010)\citenamefont{Accardi, Arleo,
  Brooks, D'Enterria, and Muccifora}}]{Accardi:2009qv}
\bibinfo{author}{\bibfnamefont{A.}~\bibnamefont{Accardi}},
  \bibinfo{author}{\bibfnamefont{F.}~\bibnamefont{Arleo}},
  \bibinfo{author}{\bibfnamefont{W.}~\bibnamefont{Brooks}},
  \bibinfo{author}{\bibfnamefont{D.}~\bibnamefont{D'Enterria}},
  \bibnamefont{and}
  \bibinfo{author}{\bibfnamefont{V.}~\bibnamefont{Muccifora}},
  \bibinfo{journal}{Riv. Nuovo Cim.} \textbf{\bibinfo{volume}{32}},
  \bibinfo{pages}{439} (\bibinfo{year}{2010}), \eprint{0907.3534}.

\bibitem[{\citenamefont{Dupr\'e}(2011)}]{Dupre:2011afa}
\bibinfo{author}{\bibfnamefont{R.}~\bibnamefont{Dupr\'e}}, Ph.D. thesis,
  \bibinfo{school}{Lyon, IPN} (\bibinfo{year}{2011}).

\bibitem[{\citenamefont{Li and Vitev}(2021)}]{Li:2020rqj}
\bibinfo{author}{\bibfnamefont{H.~T.} \bibnamefont{Li}} \bibnamefont{and}
  \bibinfo{author}{\bibfnamefont{I.}~\bibnamefont{Vitev}},
  \bibinfo{journal}{Phys. Rev. Lett.} \textbf{\bibinfo{volume}{126}},
  \bibinfo{pages}{252001} (\bibinfo{year}{2021}), \eprint{2010.05912}.

\bibitem[{\citenamefont{Li et~al.}(2022)\citenamefont{Li, Liu, and
  Vitev}}]{Li:2021gjw}
\bibinfo{author}{\bibfnamefont{H.~T.} \bibnamefont{Li}},
  \bibinfo{author}{\bibfnamefont{Z.~L.} \bibnamefont{Liu}}, \bibnamefont{and}
  \bibinfo{author}{\bibfnamefont{I.}~\bibnamefont{Vitev}},
  \bibinfo{journal}{Phys. Lett. B} \textbf{\bibinfo{volume}{827}},
  \bibinfo{pages}{137007} (\bibinfo{year}{2022}), \eprint{2108.07809}.

\bibitem[{\citenamefont{Devereaux et~al.}(2023)\citenamefont{Devereaux, Fan,
  Ke, Lee, and Moult}}]{Devereaux:2023vjz}
\bibinfo{author}{\bibfnamefont{K.}~\bibnamefont{Devereaux}},
  \bibinfo{author}{\bibfnamefont{W.}~\bibnamefont{Fan}},
  \bibinfo{author}{\bibfnamefont{W.}~\bibnamefont{Ke}},
  \bibinfo{author}{\bibfnamefont{K.}~\bibnamefont{Lee}}, \bibnamefont{and}
  \bibinfo{author}{\bibfnamefont{I.}~\bibnamefont{Moult}}
  (\bibinfo{year}{2023}), \eprint{2303.08143}.

\bibitem[{\citenamefont{Zheng et~al.}(2014)\citenamefont{Zheng, Aschenauer, and
  Lee}}]{Zheng:2014cha}
\bibinfo{author}{\bibfnamefont{L.}~\bibnamefont{Zheng}},
  \bibinfo{author}{\bibfnamefont{E.~C.} \bibnamefont{Aschenauer}},
  \bibnamefont{and} \bibinfo{author}{\bibfnamefont{J.~H.} \bibnamefont{Lee}},
  \bibinfo{journal}{Eur. Phys. J. A} \textbf{\bibinfo{volume}{50}},
  \bibinfo{pages}{189} (\bibinfo{year}{2014}), \eprint{1407.8055}.

\bibitem[{\citenamefont{Chang et~al.}(2022)\citenamefont{Chang, Aschenauer,
  Baker, Jentsch, Lee, Tu, Yin, and Zheng}}]{Chang:2022hkt}
\bibinfo{author}{\bibfnamefont{W.}~\bibnamefont{Chang}},
  \bibinfo{author}{\bibfnamefont{E.-C.} \bibnamefont{Aschenauer}},
  \bibinfo{author}{\bibfnamefont{M.~D.} \bibnamefont{Baker}},
  \bibinfo{author}{\bibfnamefont{A.}~\bibnamefont{Jentsch}},
  \bibinfo{author}{\bibfnamefont{J.-H.} \bibnamefont{Lee}},
  \bibinfo{author}{\bibfnamefont{Z.}~\bibnamefont{Tu}},
  \bibinfo{author}{\bibfnamefont{Z.}~\bibnamefont{Yin}}, \bibnamefont{and}
  \bibinfo{author}{\bibfnamefont{L.}~\bibnamefont{Zheng}}
  (\bibinfo{year}{2022}), \eprint{2204.11998}.

\bibitem[{\citenamefont{Kang et~al.}(2016)\citenamefont{Kang, Ringer, and
  Vitev}}]{Kang:2016mcy}
\bibinfo{author}{\bibfnamefont{Z.-B.} \bibnamefont{Kang}},
  \bibinfo{author}{\bibfnamefont{F.}~\bibnamefont{Ringer}}, \bibnamefont{and}
  \bibinfo{author}{\bibfnamefont{I.}~\bibnamefont{Vitev}},
  \bibinfo{journal}{JHEP} \textbf{\bibinfo{volume}{10}}, \bibinfo{pages}{125}
  (\bibinfo{year}{2016}), \eprint{1606.06732}.

\bibitem[{\citenamefont{Dai et~al.}(2016)\citenamefont{Dai, Kim, and
  Leibovich}}]{Dai:2016hzf}
\bibinfo{author}{\bibfnamefont{L.}~\bibnamefont{Dai}},
  \bibinfo{author}{\bibfnamefont{C.}~\bibnamefont{Kim}}, \bibnamefont{and}
  \bibinfo{author}{\bibfnamefont{A.~K.} \bibnamefont{Leibovich}},
  \bibinfo{journal}{Phys. Rev. D} \textbf{\bibinfo{volume}{94}},
  \bibinfo{pages}{114023} (\bibinfo{year}{2016}), \eprint{1606.07411}.

\bibitem[{\citenamefont{von Weizsacker}(1934)}]{vonWeizsacker:1934nji}
\bibinfo{author}{\bibfnamefont{C.}~\bibnamefont{von Weizsacker}},
  \bibinfo{journal}{Z. Phys.} \textbf{\bibinfo{volume}{88}},
  \bibinfo{pages}{612} (\bibinfo{year}{1934}).

\bibitem[{\citenamefont{Williams}(1934)}]{Williams:1934ad}
\bibinfo{author}{\bibfnamefont{E.}~\bibnamefont{Williams}},
  \bibinfo{journal}{Phys. Rev.} \textbf{\bibinfo{volume}{45}},
  \bibinfo{pages}{729} (\bibinfo{year}{1934}).

\bibitem[{\citenamefont{Hinderer et~al.}(2015)\citenamefont{Hinderer, Schlegel,
  and Vogelsang}}]{Hinderer:2015hra}
\bibinfo{author}{\bibfnamefont{P.}~\bibnamefont{Hinderer}},
  \bibinfo{author}{\bibfnamefont{M.}~\bibnamefont{Schlegel}}, \bibnamefont{and}
  \bibinfo{author}{\bibfnamefont{W.}~\bibnamefont{Vogelsang}},
  \bibinfo{journal}{Phys. Rev. D} \textbf{\bibinfo{volume}{92}},
  \bibinfo{pages}{014001} (\bibinfo{year}{2015}), \bibinfo{note}{[Erratum:
  Phys.Rev.D 93, 119903 (2016)]}, \eprint{1505.06415}.

\bibitem[{\citenamefont{Ovanesyan and Vitev}(2011)}]{Ovanesyan:2011xy}
\bibinfo{author}{\bibfnamefont{G.}~\bibnamefont{Ovanesyan}} \bibnamefont{and}
  \bibinfo{author}{\bibfnamefont{I.}~\bibnamefont{Vitev}},
  \bibinfo{journal}{JHEP} \textbf{\bibinfo{volume}{06}}, \bibinfo{pages}{080}
  (\bibinfo{year}{2011}), \eprint{1103.1074}.

\bibitem[{\citenamefont{Kang et~al.}(2017{\natexlab{a}})\citenamefont{Kang,
  Ringer, and Vitev}}]{Kang:2016ofv}
\bibinfo{author}{\bibfnamefont{Z.-B.} \bibnamefont{Kang}},
  \bibinfo{author}{\bibfnamefont{F.}~\bibnamefont{Ringer}}, \bibnamefont{and}
  \bibinfo{author}{\bibfnamefont{I.}~\bibnamefont{Vitev}},
  \bibinfo{journal}{JHEP} \textbf{\bibinfo{volume}{03}}, \bibinfo{pages}{146}
  (\bibinfo{year}{2017}{\natexlab{a}}), \eprint{1610.02043}.

\bibitem[{\citenamefont{Sievert and Vitev}(2018)}]{Sievert:2018imd}
\bibinfo{author}{\bibfnamefont{M.~D.} \bibnamefont{Sievert}} \bibnamefont{and}
  \bibinfo{author}{\bibfnamefont{I.}~\bibnamefont{Vitev}},
  \bibinfo{journal}{Phys. Rev.} \textbf{\bibinfo{volume}{D98}},
  \bibinfo{pages}{094010} (\bibinfo{year}{2018}), \eprint{1807.03799}.

\bibitem[{\citenamefont{Sievert et~al.}(2019)\citenamefont{Sievert, Vitev, and
  Yoon}}]{Sievert:2019cwq}
\bibinfo{author}{\bibfnamefont{M.~D.} \bibnamefont{Sievert}},
  \bibinfo{author}{\bibfnamefont{I.}~\bibnamefont{Vitev}}, \bibnamefont{and}
  \bibinfo{author}{\bibfnamefont{B.}~\bibnamefont{Yoon}},
  \bibinfo{journal}{Phys. Lett.} \textbf{\bibinfo{volume}{B795}},
  \bibinfo{pages}{502} (\bibinfo{year}{2019}), \eprint{1903.06170}.

\bibitem[{\citenamefont{Collins and Qiu}(1989)}]{Collins:1988wj}
\bibinfo{author}{\bibfnamefont{J.~C.} \bibnamefont{Collins}} \bibnamefont{and}
  \bibinfo{author}{\bibfnamefont{J.-w.} \bibnamefont{Qiu}},
  \bibinfo{journal}{Phys. Rev. D} \textbf{\bibinfo{volume}{39}},
  \bibinfo{pages}{1398} (\bibinfo{year}{1989}).

\bibitem[{\citenamefont{Chien et~al.}(2016)\citenamefont{Chien, Emerman, Kang,
  Ovanesyan, and Vitev}}]{Chien:2015vja}
\bibinfo{author}{\bibfnamefont{Y.-T.} \bibnamefont{Chien}},
  \bibinfo{author}{\bibfnamefont{A.}~\bibnamefont{Emerman}},
  \bibinfo{author}{\bibfnamefont{Z.-B.} \bibnamefont{Kang}},
  \bibinfo{author}{\bibfnamefont{G.}~\bibnamefont{Ovanesyan}},
  \bibnamefont{and} \bibinfo{author}{\bibfnamefont{I.}~\bibnamefont{Vitev}},
  \bibinfo{journal}{Phys. Rev.} \textbf{\bibinfo{volume}{D93}},
  \bibinfo{pages}{074030} (\bibinfo{year}{2016}), \eprint{1509.02936}.

\bibitem[{\citenamefont{Altarelli and Parisi}(1977)}]{Altarelli:1977zs}
\bibinfo{author}{\bibfnamefont{G.}~\bibnamefont{Altarelli}} \bibnamefont{and}
  \bibinfo{author}{\bibfnamefont{G.}~\bibnamefont{Parisi}},
  \bibinfo{journal}{Nucl. Phys. B} \textbf{\bibinfo{volume}{126}},
  \bibinfo{pages}{298} (\bibinfo{year}{1977}).

\bibitem[{\citenamefont{Ke and Vitev}(2022)}]{Ke:2022gkq}
\bibinfo{author}{\bibfnamefont{W.}~\bibnamefont{Ke}} \bibnamefont{and}
  \bibinfo{author}{\bibfnamefont{I.}~\bibnamefont{Vitev}}
  (\bibinfo{year}{2022}), \eprint{2204.00634}.

\bibitem[{\citenamefont{Salam and Rojo}(2009)}]{Salam:2008qg}
\bibinfo{author}{\bibfnamefont{G.~P.} \bibnamefont{Salam}} \bibnamefont{and}
  \bibinfo{author}{\bibfnamefont{J.}~\bibnamefont{Rojo}},
  \bibinfo{journal}{Comput. Phys. Commun.} \textbf{\bibinfo{volume}{180}},
  \bibinfo{pages}{120} (\bibinfo{year}{2009}), \eprint{0804.3755}.

\bibitem[{\citenamefont{Kang et~al.}(2017{\natexlab{b}})\citenamefont{Kang,
  Ringer, and Vitev}}]{Kang:2017frl}
\bibinfo{author}{\bibfnamefont{Z.-B.} \bibnamefont{Kang}},
  \bibinfo{author}{\bibfnamefont{F.}~\bibnamefont{Ringer}}, \bibnamefont{and}
  \bibinfo{author}{\bibfnamefont{I.}~\bibnamefont{Vitev}},
  \bibinfo{journal}{Phys. Lett.} \textbf{\bibinfo{volume}{B769}},
  \bibinfo{pages}{242} (\bibinfo{year}{2017}{\natexlab{b}}),
  \eprint{1701.05839}.

\bibitem[{\citenamefont{Li and Vitev}(2019)}]{Li:2018xuv}
\bibinfo{author}{\bibfnamefont{H.~T.} \bibnamefont{Li}} \bibnamefont{and}
  \bibinfo{author}{\bibfnamefont{I.}~\bibnamefont{Vitev}},
  \bibinfo{journal}{JHEP} \textbf{\bibinfo{volume}{07}}, \bibinfo{pages}{148}
  (\bibinfo{year}{2019}), \eprint{1811.07905}.

\bibitem[{\citenamefont{Dulat et~al.}(2016)\citenamefont{Dulat, Hou, Gao,
  Guzzi, Huston, Nadolsky, Pumplin, Schmidt, Stump, and Yuan}}]{Dulat:2015mca}
\bibinfo{author}{\bibfnamefont{S.}~\bibnamefont{Dulat}},
  \bibinfo{author}{\bibfnamefont{T.-J.} \bibnamefont{Hou}},
  \bibinfo{author}{\bibfnamefont{J.}~\bibnamefont{Gao}},
  \bibinfo{author}{\bibfnamefont{M.}~\bibnamefont{Guzzi}},
  \bibinfo{author}{\bibfnamefont{J.}~\bibnamefont{Huston}},
  \bibinfo{author}{\bibfnamefont{P.}~\bibnamefont{Nadolsky}},
  \bibinfo{author}{\bibfnamefont{J.}~\bibnamefont{Pumplin}},
  \bibinfo{author}{\bibfnamefont{C.}~\bibnamefont{Schmidt}},
  \bibinfo{author}{\bibfnamefont{D.}~\bibnamefont{Stump}}, \bibnamefont{and}
  \bibinfo{author}{\bibfnamefont{C.~P.} \bibnamefont{Yuan}},
  \bibinfo{journal}{Phys. Rev.} \textbf{\bibinfo{volume}{D93}},
  \bibinfo{pages}{033006} (\bibinfo{year}{2016}), \eprint{1506.07443}.

\bibitem[{\citenamefont{Buckley et~al.}(2015)\citenamefont{Buckley, Ferrando,
  Lloyd, Nordstrom, Page, Ruefenacht, Schoenherr, and Watt}}]{Buckley:2014ana}
\bibinfo{author}{\bibfnamefont{A.}~\bibnamefont{Buckley}},
  \bibinfo{author}{\bibfnamefont{J.}~\bibnamefont{Ferrando}},
  \bibinfo{author}{\bibfnamefont{S.}~\bibnamefont{Lloyd}},
  \bibinfo{author}{\bibfnamefont{K.}~\bibnamefont{Nordstrom}},
  \bibinfo{author}{\bibfnamefont{B.}~\bibnamefont{Page}},
  \bibinfo{author}{\bibfnamefont{M.}~\bibnamefont{Ruefenacht}},
  \bibinfo{author}{\bibfnamefont{M.}~\bibnamefont{Schoenherr}},
  \bibnamefont{and} \bibinfo{author}{\bibfnamefont{G.}~\bibnamefont{Watt}},
  \bibinfo{journal}{Eur. Phys. J.} \textbf{\bibinfo{volume}{C75}},
  \bibinfo{pages}{132} (\bibinfo{year}{2015}), \eprint{1412.7420}.

\bibitem[{\citenamefont{Hirai et~al.}(2007)\citenamefont{Hirai, Kumano, Nagai,
  and Sudoh}}]{Hirai:2007cx}
\bibinfo{author}{\bibfnamefont{M.}~\bibnamefont{Hirai}},
  \bibinfo{author}{\bibfnamefont{S.}~\bibnamefont{Kumano}},
  \bibinfo{author}{\bibfnamefont{T.-H.} \bibnamefont{Nagai}}, \bibnamefont{and}
  \bibinfo{author}{\bibfnamefont{K.}~\bibnamefont{Sudoh}},
  \bibinfo{journal}{Phys. Rev. D} \textbf{\bibinfo{volume}{75}},
  \bibinfo{pages}{094009} (\bibinfo{year}{2007}), \eprint{hep-ph/0702250}.

\bibitem[{\citenamefont{Braaten et~al.}(1995)\citenamefont{Braaten, Cheung,
  Fleming, and Yuan}}]{Braaten:1994bz}
\bibinfo{author}{\bibfnamefont{E.}~\bibnamefont{Braaten}},
  \bibinfo{author}{\bibfnamefont{K.-m.} \bibnamefont{Cheung}},
  \bibinfo{author}{\bibfnamefont{S.}~\bibnamefont{Fleming}}, \bibnamefont{and}
  \bibinfo{author}{\bibfnamefont{T.~C.} \bibnamefont{Yuan}},
  \bibinfo{journal}{Phys. Rev. D} \textbf{\bibinfo{volume}{51}},
  \bibinfo{pages}{4819} (\bibinfo{year}{1995}), \eprint{hep-ph/9409316}.

\bibitem[{\citenamefont{Cheung and Yuan}(1996)}]{Cheung:1995ye}
\bibinfo{author}{\bibfnamefont{K.-m.} \bibnamefont{Cheung}} \bibnamefont{and}
  \bibinfo{author}{\bibfnamefont{T.~C.} \bibnamefont{Yuan}},
  \bibinfo{journal}{Phys. Rev. D} \textbf{\bibinfo{volume}{53}},
  \bibinfo{pages}{1232} (\bibinfo{year}{1996}), \eprint{hep-ph/9502250}.

\end{thebibliography}

\end{document}